\documentclass[12pt]{article}
\pdfoutput=1
\usepackage{amsmath,amssymb,amsfonts,amsthm,bm,bbm,cancel,wasysym}
\usepackage{epsfig,graphics,graphicx,epstopdf}
\usepackage{array,booktabs,colortbl,colordvi,multirow}
\usepackage{colordvi,color,xcolor}
\usepackage{hyperref}
\usepackage{rotating}

\usepackage{verbatim}
\usepackage{cite}
\usepackage{subfig}
\usepackage{setspace}
\usepackage{url}
\usepackage[percent]{overpic}
\usepackage{slashed}
\usepackage{authblk}
\usepackage{xspace}
\usepackage{fullpage}

\def\ie{{\it i.e.}}
\def\eg{{\it e.g.}}

\def\to{\rightarrow}

\newskip\zatskip \zatskip=0pt plus0pt minus0pt
\def\matth{\mathsurround=0pt}
\def\lsim{\mathrel{\mathpalette\atversim<}}
\def\gsim{\mathrel{\mathpalette\atversim>}}
\def\atversim#1#2{\lower0.7ex\vbox{\baselineskip\zatskip\lineskip\zatskip
  \lineskiplimit 0pt\ialign{$\matth#1\hfil##\hfil$\crcr#2\crcr\sim\crcr}}}

\newcommand{\emailto}[1]{E-mail: \href{mailto:#1}{\protect \nolinkurl{#1}}} 

\begin{document}


\begin{flushright}
SLAC-PUB-16856 \\
\today
\end{flushright}
\vspace*{5mm}

\renewcommand{\thefootnote}{\fnsymbol{footnote}}
\setcounter{footnote}{1}

\begin{center}

{\Large {\bf Gravity-Mediated Dark Matter Annihilation in}\\
{\bf the Randall-Sundrum Model}}\\

\vspace*{0.75cm}
{\bf T. D. Rueter${}^{1,2}$}~\footnote{\emailto{tdr38@stanford.edu}}
{\bf T. G. Rizzo${}^2$}~\footnote{\emailto{rizzo@slac.stanford.edu}} and
{\bf J. L. Hewett${}^2$}~\footnote{\emailto{hewett@slac.stanford.edu}},

\vspace{0.5cm}

${}^1${Stanford University, Stanford, CA, USA}

${}^2${SLAC National Accelerator Laboratory, Stanford University, Menlo Park, CA, USA}

\end{center}
\vspace{.5cm}

\begin{abstract}
 
\noindent

Observational evidence for dark matter stems from its gravitational interactions, and as of yet there has been no evidence for 
dark matter interacting via other means. We examine models where dark matter interactions are purely gravitational in a 
Randall-Sundrum background. In particular, the Kaluza-Klein tower of gravitons which result from the warped fifth dimension can 
provide viable annihilation channels into Standard Model final states, and we find that we can achieve values of the annihilation 
cross section, $\left< \sigma v \right>$, which are consistent with the observed relic abundance in the case of spin-1 dark 
matter. We examine constraints on these models employing both the current photon line and continuum indirect dark matter 
searches, and assess the prospects of hunting for the signals of such models in future direct and indirect detection experiments.

\end{abstract}

\renewcommand{\thefootnote}{\arabic{footnote}}
\setcounter{footnote}{0}
\thispagestyle{empty}
\vfill
\newpage
\setcounter{page}{1}



\section{Introduction}
\label{Sections/Section_1_Intro}

Observations indicate that roughly 85\% of the matter in the universe is non-luminous, with this conclusion being drawn from the apparent gravitational effects of cold dark matter particles.  The nature of this dark matter (DM), and its potential interactions with the
Standard Model (SM), remain mysterious.  A widely studied DM candidate is
the thermal relic weakly interacting massive particle (WIMP) scenario, where the observed relic density implies that the DM has a mass in the 
GeV-TeV range and possibly couples with weak interaction strength (or less) to the SM fields directly or
through an additional mediator field \cite{steigman1985cosmological, bertone2005particle}. If the DM is a SM singlet field, a mediator is
a necessary ingredient in order to generate non-gravitational signatures, and the DM itself may only be a small part of a larger Dark Sector.  Despite numerous attempts, a non-gravitational DM signal has yet to be observed and the WIMP-like scenario is already reasonably constrained by both direct \cite{Akerib:2016vxi, Tan:2016zwf} and 
indirect \cite{Ackermann:2015lka, Abramowski:2013ax, Abdalla:2016olq, Lefranc:2016fgn, Ackermann:2015zua, Drlica-Wagner:2015xua, Ahnen:2016qkx} searches, as well as by complimentary DM and mediator searches at low energy accelerators \cite{Alexander:2016aln} and the LHC \cite{LHC13}.   

In this paper, we examine DM in theories with additional spatial dimensions where the DM itself has purely gravitational interactions.  We work within 
the framework of the warped extra dimension scenario of Randall and Sundrum (RS) \cite{Randall:1999ee} where  DM
communication with the SM may then occur via a Kaluza-Klein graviton mediator present in the model.  The RS model is an attractive
scenario, presenting well-known solutions to the gauge hierarchy and the fermion mass hierarchy problems.  In this framework 
the gauge hierarchy is generated via an exponential warp-factor appearing in the 5-dimensional metric, with the Higgs field, together with its vev,
being located on or near the so-called TeV-scale brane.   Mass scales are thus reduced by this exponential factor in comparison to the Planck scale.
In addition, the suitable placement of the SM gauge and fermion fields in the 5-dimensional bulk  \cite{Gherghetta:2000qt}, 
allows for the reduction of the fermion mass
hierarchy to that of selecting an appropriate set of $\cal{O}$(1) numbers. Of course this RS picture is not without its own problems: 
It faces some rather severe constraints from ($i$) precision electroweak measurements \cite{Carena:2004zn}, ($ii$) the existence of 
tree-level Flavor-Changing Neutral Currents \cite{Huber:2000ie}, as well as ($iii$) direct searches for the predicted Kaluza-Klein (KK) excitations of the graviton and SM gauge fields at the LHC\cite{LHC13}.  Together, these constraints generally force the KK scale near the $\sim 10$ TeV range even when there is a custodial gauge symmetry present \cite{Agashe:2003zs}.{\footnote {See, however, \cite{us} for a 
possible way to evade these significant issues and reduce the KK gauge boson masses to $\sim 1$ TeV.}}  

The RS scenario naturally realizes the interesting possibility that the DM communicates with the SM via purely gravitational interactions, the only 
interactions that we know for sure that DM possesses, utilizing the KK graviton excitations as mediators between the DM and the SM.
This concept was pioneered in \cite{Han:2015cty, Lee:2013bua, Lee:2014caa} and has been adapted for use in the form of a Simplified Model in  \cite{Kraml:2017atm} along the 
lines suggested by the LHC Dark Matter Working Group  \cite{Abdallah:2015ter}.   Here, we explore this enticing possibility in more detail 
by employing a number of specific benchmark scenarios which make full use of the RS model building features of varying
localizations for the SM gauge and fermion fields, brane localized kinetic terms \cite{georgi2001brane}, as well as differing
graviton KK mass scales.  In particular, $(i)$ we provide an exact treatment of the graviton exchange process leading to the velocity weighted thermal DM annihilation cross section, without employing a velocity expansion.   As we will see below, the velocity expansion approximation is questionable 
near KK resonances and such regions are found to be of particular importance in our analysis.
($ii$) We consider in detail the various possibilities of the DM spin, spin-$0, 1/2$ and 1, and determine the DM velocity dependence for 
all potential SM annihilation channels.  We demonstrate that the success of this scenario requires the DM to be spin-1 and have a mass  
$\gsim 350$ GeV to avoid velocity suppression of the annihilation cross section. ($iii$)  We include the contributions of multiple graviton KK excitations; this is found to be important since the behavior of the thermal cross section near these additional peaks opens up new successful parameter space regions. However, we observe that the relative behavior of the KK resonance regions vary in detail between the 
different  benchmark scenarios. ($iv$) We examine the direction and indirect detection constraints in detail. As we will show below, indirect detection
searches for both antiprotons and gamma rays are found to potentially constrain interesting parameter space regions, while there is no observable
signature in direct detection experiments. 

The organization of this paper is as follows:  After our Introduction in the present Section, we provide an overview of the basics of the RS framework in
Section 2 where our benchmark models are also presented. Section 3 contains our analysis of the thermal relic abundance via graviton KK exchange 
for DM particles of various spins, as well as a discussion of the constraints that arise from direct and indirect DM searches. A discussion and our conclusions can be found in Section 4. The Appendix summarizes the formulae for 
the KK graviton partial widths, as well as the differential cross sections for DM annihilation of various spins.



\section {Randall-Sundrum Model Framework}
\label{section:Model}

In this Section we discuss the general model framework that we follow in the analysis below.   The RS scenario is based on a slice of AdS$_5$
spacetime bounded by two 4-dimensional Minkowski branes.  The fifth dimension, denoted by $y=r_c\phi$, is bounded at either end by
the UV-, or Planck, brane located at $y=0$ and the IR-, or TeV, brane sitting at $y=\pi r_c$.  The associated metric is 
$ds^2=e^{-2ky}\eta_{\mu\nu}dx^\mu dx^\nu - dy^2$. The exponential represents the warp factor (which generates the gauge hierarchy), 
while the parameter $k$ governs the degree
of curvature of the AdS$_5$ space.  The relation $\overline{M}^2_{Pl}=M_5^3/k$ is thus derived from the 5-D action.
The scale of physical phenomena on the IR-brane is given by $\Lambda_\pi\equiv \overline{M}_{Pl}
e^{-kr_c\pi}$ with $\Lambda_\pi\sim$ few TeV implying $kr_c\sim 11-12$.  In the simplest picture, the SM fields reside on the IR-brane with only
gravity propagating in the $5^{th}$ dimension.  The mass of the $n^{th}$ KK graviton excitation is $m_n^G=x_n^Gk\Lambda_\pi/\overline{M}_{Pl}$,
with $x_n^G$ being the roots of the $J_1$ Bessel function.  The coupling strength of the graviton KK states to the SM fields is $\Lambda_\pi^{-1}$ 
\cite{DHR}.

The small size of the additional dimension allows for the SM fields to propagate in the bulk.  This scenario has attractive model-building
features such as the potential to explain the fermion mass hierarchy.  In this case the gauge bosons also have KK excitations with their
masses being governed by the roots of the equation
\begin{equation}
 J_1(x_n^A)+x_n^AJ_1'(x_n^A) + \alpha_n[Y_1(x_n^A)+x_n^AY_1'(x_n^A)]=0 \,,
 \end{equation}
 with the numerical coefficients $\alpha_n$ being determined by the boundary conditions on the UV-brane.  The gauge KK masses are then
 $m_n^A=x_n^Ak\Lambda_\pi/\overline{M}_{Pl}$.  When the fermions also reside in the bulk, additional parameters ($m_i=k\nu_i$
 or $=-kc_i$) are introduced corresponding to the bulk fermion masses.  These parameters determine the localization of the fermion
 wavefunction.  As the Higgs boson is kept on or near the IR-brane (to address the gauge hierarchy), a consistent framework emerges
 if light flavor fermions are localized near the UV-brane and the third generation quarks are localized on or near the IR-brane.  The addition
 of brane localized kinetic terms (BLKTs), which arise naturally from quantum effects \cite{georgi2001brane}, modify the spectrum and couplings of the
 graviton, gauge, and fermion KK states \cite{DHR}.

To be specific, we 
consider five benchmark points in the RS parameter space that differ in the localizations of the various SM and DM 
particles and in the masses of the lowest lying KK gravitons. For all of these benchmarks the Higgs field and the 
associated Goldstone bosons, \ie, essentially the longitudinal components of the massive gauge bosons that result from spontaneous 
symmetry breaking, reside on the IR-brane. To be general, we will in some cases, also allow for the presence of brane localized 
kinetic terms (BLKTs) for the various bulk fields so that, in the notation of Ref. {\cite{DHR}}, the 5-D gravitational action will 
be given by
\begin{eqnarray}
 S_G = & \frac{M_5^3}{4} & \int d^4 x \int r_c d\phi \, \sqrt {-G} \,
\left\{R^{(5)} + [2\gamma_0/kr_c ~\, \delta(\phi) \right. \nonumber\\ 
& + &
\left. 2\gamma_\pi/kr_c ~\, \delta(\phi - \pi)] \, R^{(4)} + \ldots \right\}\,,
\end{eqnarray}
where $\gamma_{0,\pi}$ are the corresponding dimensionless UV- and IR-brane terms, respectively. Similarly, for 
the SM and possible DM gauge fields we have
\begin{eqnarray}
 S_V= & \frac{-1}{4} & \int d^4 x \int r_c d\phi \, \sqrt {-G} \,
\left\{F_{AB}F^{AB} + [2\delta_0/kr_c ~\, \delta(\phi) \right. \nonumber\\ 
& + &
\left. 2\delta_\pi/kr_c ~\, \delta(\phi - \pi)] \, F_{\mu\nu}F^{\mu\nu} + \ldots \right\}\,,
\end{eqnarray}
where $\delta_{0,\pi}$ are the UV- and IR-brane terms. We assume for simplicity that the gauge fields 
corresponding to the three SM gauge groups all have common BLKT values $\delta_0$ and $\delta_\pi$, and generically $\delta_0 \neq \delta_\pi$. If the DM is identified as a spin-1 gauge bulk field, 
this assumption will apply to the DM field as well. For the various fermions in the bulk {\cite{Carena:2004zn}} we have similarly 
\begin{eqnarray}
 S_F= &\frac{}{} &\int d^4 x \int r_c d\phi \, \sqrt{-G} \, 
\left\{\frac{-i}{2}\bar\Psi \Gamma^A \partial_A \Psi + [2\tau_0/kr_c ~\, \delta(\phi) \right. \nonumber\\ 
& + & 
\left. 2\tau_\pi/kr_c ~\, \delta(\phi - \pi)] \, \frac{-i}{2}\bar\Psi \gamma^\mu \partial_\mu \Psi +h.c. \right. \\
& - &
\left. sgn(\phi) ~m_{f_\Psi} \bar \Psi \Psi  \frac{}{} \right\} \nonumber\,,
\end{eqnarray}
where $\tau_{0,\pi}$ are the corresponding UV- and IR-brane terms which we again assume take on identical values 
for all SM fermion representations for simplicity. As usual, each of the various fermion representations take on distinct 
discrete values for the bulk mass parameter, $m_f=k\nu_f$, which will peak their wavefunctions towards either the IR- or UV- 
branes. These values are predominantly determined by replicating the fermion mass hierarchy.

In the original RS model where all the SM fields reside on the TeV brane, and in the absence of graviton BLKTs, the 
SM fields couple to the full set  
of graviton KK excitations with universal strength, $\Lambda_\pi^{-1}$. Here we scale the 
couplings of each field to this quantity, so the effect of both the field localizations and the various BLKTs can be easily 
tracked. For example, the existence of graviton BLKTs rescales the KK graviton tower couplings to the fields localized on the TeV brane by factors of (here the index $n$ labels the state in the graviton KK tower)   
\begin{equation}
\lambda_n\equiv \Big[ {{1+2\gamma_0} \over {1+(x^G_n\gamma_\pi)^2-2\gamma_\pi}} \Big]^{1/2}\,.
\label{lambdaeq}
\end{equation}
This \lq correction' factor clearly takes on a value of unity in the absence of graviton BLKTs. Recall that the graviton KK masses themselves are also influenced by the presence of BLKTs and are given by $m^G_n=x^G_nk\epsilon=
x^G_n k\Lambda_\pi/{\overline M}_{Pl}$, with $x^G_n$ being the roots of the transcendental equation
\begin{equation}
J_1(x^G_n) - \gamma_\pi x^G_n J_2(x^G_n)=0\,,
\label{gravroots}
\end{equation}
where $J_{1,2}$ are the usual Bessel functions of the first kind. 

For the case of bulk gauge fields, the graviton couplings to the 5-D Yang-Mills kinetic term in the presence of BLKTs are {\it further} rescaled by additional factors of  
\begin{equation}
\label{del}
\delta_n = {{2(1-J_0(x^G_n))+(\delta_\pi-\gamma_\pi)(x^G_n)^2 J_2(x^G_n)}\over {(\pi kr_c+\delta_\pi+\delta_0) (x^G_n)^2|J_2(x^G_n))|}} \,.
\end{equation}
The graviton KK coupling strengths, in units of $\Lambda_\pi^{-1}$, are given by the product $\lambda_n \delta_n$. We briefly note that while the graviton KK states couple to the 5-D Yang-Mills kinetic Lagrangian for bulk gauge fields, this is sometimes referred to in the literature as a coupling to the transverse polarizations of the gauge fields in the bulk. This is certainly the case for massless gauge bosons, but for massive gauge bosons where the spontaneous symmetry breaking occurs on the IR-brane, the kinetic part of the coupling of the longitudinal polarization to the $n^{th}$ graviton KK mode also arises from the 5-D Yang-Mills kinetic term, and thus is rescaled by a factor of $\delta_n$. We emphasize that the Higgs-gauge interactions on the IR-brane, including spontaneous symmetry breaking, are not rescaled by the $\delta_n$ factor.

Correspondingly, for the case where the SM fermions, $f_{L,R}$, reside in the bulk, in addition to the overall factor of $\lambda_n$ above, their 
couplings to the graviton KK tower fields are further rescaled by factors of $c_{L,Rn}^f$. These are the coupling strengths of left-handed and right-handed bulk fermions $f$ to the $n^{th}$ graviton KK state, and are determined through numerical 
integration over the appropriate 5-D wavefunctions. Clearly, the factors $c^f_{L,Rn}$ will be $\nu_{f_{L,R}}$-dependent. The set of Feynman rules that are required to compute the DM annihilation cross section are given explicitly in the next Section.

\begin{figure}[htbp]
\vspace*{-1.5cm}
\begin{center}
\includegraphics[scale=0.50,angle=90]{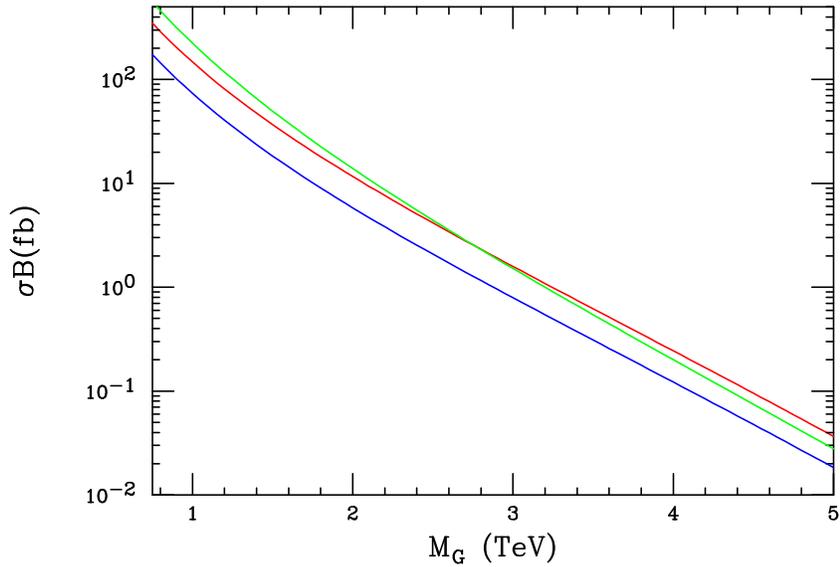}
\end{center}
\vspace*{-1.50cm}
\caption{Production cross sections for the lightest KK graviton production at the 13 TeV LHC for several of our benchmark models.
For the cases where the SM is constrained to the TeV brane the red(blue) curve is weighted by the branching fraction into the 
$\gamma\gamma(\ell^+\ell^-)$ final states with production occurring through both $gg$ and $q\bar q$ annihilation. In the GW benchmark case (green), the 
result includes the branching fraction into diphotons and only has contributions from the $gg$ production channel since there is essentially 
no coupling to light quarks (or leptons) in this case. For the MOR benchmark, the result is similar to that for GW apart from an overall 
normalization factor of $\simeq$ 0.03.}
\label{lhc}
\end{figure}

We now define our five benchmark points.  These points have been chosen in order to provide distinct examples of the various RS
model features in order to illustrate the potential variation in the results.  
The first 3 benchmarks are rather simple: in these cases we assume that all of the SM fields, as well as the DM, are constrained to lie on the IR-brane; we denote these benchmark points as Brane models. These points 
differ only in the assumed value of the mass of the lightest graviton KK state in each case, which we take to be 750, 1500 or 
3000 GeV, respectively. Since in these cases the only free parameter in the model is $\Lambda_\pi$, we must choose this parameter such 
that we are not in conflict with the 13 TeV LHC searches for graviton production and decay in all channels, with the
most relevant being: $G_1 \to jj, \gamma \gamma $ and $\ell^+ \ell^-${\cite {LHC13}}.{\footnote {Of course we will ensure 
that these constraints are also satisfied for the other benchmark models as well.}} Fig.~\ref{lhc} shows the production cross 
sections for the first graviton KK state at the 13 TeV LHC for the relevant diphoton and dilepton final state from which 
the $\Lambda_\pi$ limits can be extracted. The dijet mode is always found to yield an inferior constraint, since for SM 
fields on the IR-brane one finds $\sigma_{jj}/\sigma_{\gamma\gamma} \simeq 14$, which is not enough to overcome the much larger SM background for dijet production. To lie safely below the current constraints for the lightest KK graviton masses assumed above, we choose $\Lambda_\pi = 135(90,30)$ TeV, respectively, for these 3 benchmark models. Note that these benchmark points do not address the gauge hierarchy.

The other two benchmark models are more complex. In the first case (hereafter referred to as MOR), the details of the model 
are provided in Ref.{\cite {Hewett:2016omf}}. In this case the third generation 
quarks are confined to the IR-brane while all other SM fermions are localized sufficiently close to the UV-brane 
so that their couplings to the graviton KK modes can be safely ignored. Gauge fields are placed in the bulk. Here one 
finds, \eg, that $\lambda_1 \simeq 0.125$, $\delta_1=0.5$ with $\Lambda_\pi =6$ TeV, which is chosen to satisfy the LHC search 
constraints above given an assumed mass of the lightest KK graviton of 750 GeV. The values of the BLKT parameters are chosen to be
$\gamma_\pi=-7.652$, $\delta_{0,\pi}=-10$ and $1+2\gamma_0=25$. Given 
these inputs the values of $\lambda_n,\delta_n$ can be easily calculated. Note that the MOR model satisfies the original motivation of RS models in explaining the gauge hierarchy.  It was originally motivated by the bygone 750 GeV diphoton excess, but remains in agreement with the data for this value of $\Lambda_\pi$.

The final benchmark point (hereafter referred to as GW) is significantly more complex {\cite {us}}. In this case all SM fermions 
are localized at various places in the bulk and the gauge, fermion and graviton BLKTs are assumed to be universal, \ie, 
$\tau_\pi=\gamma_\pi=\delta_\pi$ for the IR-brane and similarly for the UV BLKTs. $\nu_{t_R}=-0.5$ and $\nu_{b_R} \simeq -0.6$ are assumed. 
The values of $\tau_{\pi,0}$ and the other $\nu_f$ parameters are then floated to obtain the observed fermion masses, the 
correct structure and values for the CKM matrix elements and a sufficiently strong suppression of the couplings of the SM 
zero-mode fermions to the KK gauge tower fields. The avoidance of FCNCs and the absence of potentially large gauge couplings 
to the KK graviton tower are also employed as additional constraints. Given these conditions, rather unique ranges for the 
values of the remaining parameters are obtained in this scenario; the value of $\lambda_1^{-1}\Lambda_\pi= 28$ TeV with 
a lightest KK graviton mass of 3 TeV will be employed in the numerical study of this benchmark below although other values 
are possible. For more details, see Ref.{\cite {us}}.

\begin{figure}[htbp]
\vspace*{-2.0cm}
\begin{center}
\includegraphics[scale=0.50,angle=90]{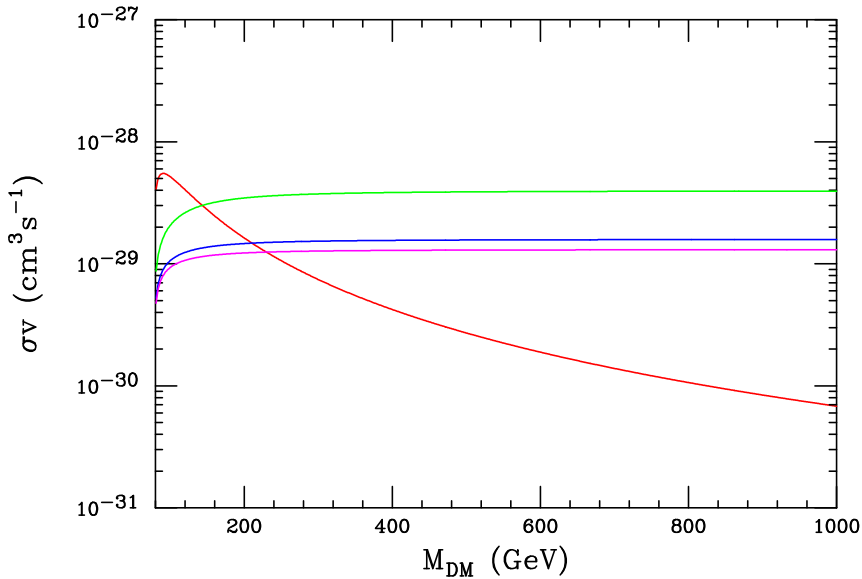}
\end{center}
\vspace*{-1.50cm}
\caption{Higgs-mediated pair annihilation cross section scaled by the relative DM velocity as a function of the DM mass; note that this cross 
section is not thermally averaged. The red curve corresponds to the case of scalar dark matter, $S$, for the specific process $SS\to 
h\to W^+W^-$, with $\tilde \lambda=10^{-2}$ (as discussed in the text). The 
green (blue, magenta) curves are for the analogous vector DM annihilation channel assuming an effective  coupling of $10^{-2}$ 
as described in the text. Here the heavier scalar mass is assumed to be $3(5,7)m_{DM}$, respectively.}
\label{demo}
\end{figure}

We next discuss the properties of the DM field. To be explicit, we will generally allow the DM to be either a real scalar, a neutral, vector-like 
Dirac fermion, or a new neutral massive gauge field, where in all cases it will be assumed to be a SM singlet. For all these 
possibilities we will assume that the DM states annihilate to SM fields solely (or dominantly) via their KK graviton 
interactions, in line with the premise of this paper. However, this introduces some potential difficulties when building a realistic model, which we now discuss.
 For example, when we introduce a new real scalar field $S$ as the DM it is well-known that the scalar portal mechanism may be 
operable{\cite {scalarportals}}. In this case, nothing forbids the $S$ from interacting with the SM fields via an induced 
trilinear coupling to the Higgs of the form $\sim \tilde \lambda v_H Sh^2$ which results from SSB in the SM. For example, 
if $m_S>M_W$, the DM annihilation process $SS\to h^* \to W^+W^-$ can occur in addition to annihilation 
through the graviton KK channels and may even dominate. However, we find that if $\tilde \lambda \lsim 10^{-2}$, then these 
SM-mediated processes have sufficiently small rates so that $S$ can never achieve the thermal relic density target via 
the portal mechanism; this is illustrated in Fig.~\ref{demo}.  We assume this small coupling in the analysis below, and ignore 
this potential annihilation contribution. For all the benchmark points considered here we take the DM singlet scalar to be localized on the TeV brane,
as is the case for the SM Higgs. 

In the case of vector DM (which we assume is a new $U(1)_D$ gauge field $V$), the situation is a bit more complex. First, 
we assume that the mass of this gauge field is generated via its coupling to a real scalar field, $S$, localized on the 
IR-brane, which gets a TeV-scale vev, $v_S$, via the usual Higgs mechanism. Necessarily, we must have $m_S>m_V$ so 
that $V$ is the lightest new state and can be identified as the DM. In this case, since $S$ gets a vev, $S$ must mix with the 
usual SM Higgs field through a phenomenologically small (constrained by LHC measurements \cite{LHCconst}) angle, $\theta$, forming the mass eigenstates $h_{1,2}$. Since a 
$VVS$ coupling is generated by the non-zero vev $v_S$, the DM $V$ can pair annihilate to SM fields via, \eg, the process 
$VV\to h_{1,2}\to W^+W^-$, which is controlled by the overall effective coupling $(g_D/g) s_\theta c_\theta$ where $g_D[g]$ is 
the value of the $U(1)[SU(2)_L]$ gauge coupling.  As shown in Fig.~\ref{demo}, taking $(g_D/g) s_\theta c_\theta \lsim 
10^{-2}$ with various assumed values of the mass of $h_2$, we see that this mechanism does not lead to a large enough 
cross section to generate 
the observed relic density. We assume that this coupling is sufficiently suppressed in the analysis below, 
so that we can safely ignore this process in our further discussion. Similar arguments to these can be made in order to avoid a 
substantial corresponding contribution to DM pair annihilation arising from a gauge portal{\cite {vectorportals}}.     
For example, we assumed (i) that there are no states in the model that share both SM and new $U(1)_D$ couplings at lowest order in 
the absence of mixing among the matter fields, or that (ii) there exists a $Z_2$ symmetry under which V is odd so that kinetic mixing is absent. 

The remaining property to be determined in our benchmark models of gravitationally interacting DM is the localization of the DM itself within the extra dimension. In the first three cases where the SM is IR-brane localized, we localize the DM there as well. In the case of scalar (vector) DM, its coupling to the KK graviton
states is then the same as that for the Higgs (SM $Z$) boson apart from the value of the DM mass. 

When the DM is a fermion, it also is assumed to couple to the KK gravitons in the same way as does any other similarly localized 
SM fermion, apart from its mass. The DM localization assignments are more complex in both the MOR and GW 
benchmark scenarios. In the MOR case, for scalar (vector) DM, the $S (V)$ field is localized to the IR-brane as is the SM 
Higgs. In the case of fermionic DM for the MOR benchmark, the DM field is localized to the TeV brane and couples in a 
manner analogous to the third generation of quarks. For the GW model, fermionic DM lies in the bulk and couples in a 
manner similar to the right-handed top quark. For vector DM in this case, $V$ is assumed to behave in a manner similar 
to the SM $Z$ boson apart from its mass. 

In all cases we treat the DM mass as a free parameter except for the requirement  
that $m_{DM} < m_{G^1}$, which we assume for simplicity, thus avoiding DM annihilation into graviton KK states.{\footnote {We will make this 
assumption for all DM spin assignments in all benchmark cases independently of where the DM is localized.}} Once above this threshold, the 
process obtains contributions from DM exchanges in both the $t$- and $u$-channels, from a 4-pt coupling, as well as 
via multiple $s$-channel KK graviton exchanges. These $s$-channel terms will clearly dominate in the regions near graviton 
KK resonances through triple graviton coupling processes such as $G_3 \to 2G_1$.{\footnote {Recall that the graviton KK spectrum satisfies the ordering 
$m_{G_2} < 2m_{G_1} < m_{G_3}$.}} These types of decays (in the absence of 
graviton BLKTs) were considered in Ref.{\cite {Davoudiasl:2001uj}} and the analysis presented there can be straightforwardly 
generalized to cover this more complex situation.  To estimate the size of such contributions, we consider the simplest situation, 
corresponding to our first three benchmark models, where all SM fields are constrained to lie on the IR-brane. For 
these benchmarks we find that $B(G_3 \to 2G_1)\simeq 15.2\%$, indicating that this would be an important, though not dominant, final 
state and that the overall DM annihilation cross section in this mass range would not be much influenced by the presence of this 
contribution. Note that in the mass range $2m_{DM}< m_{G_3}$, phase space arguments suggest that this $2G_1$ final state will be even 
less important. Thus we conclude that our results would not be significantly altered if our assumption of $m_{DM} \leq m_{G_1}$ is violated. 

We now turn to a discussion of DM annihilation to the various SM fields for each of these benchmark scenarios.



\section{Analysis}
\label{section:analysis}

In this section we describe the computation of the thermal relic abundance, as well as the cross sections for direct and indirect detection for gravitationally mediated DM interactions. 

\subsection{Thermal Relic Abundance Calculation}

Before proceeding with our thermal relic calculation, we note that the Randall-Sundrum geometry, when stabilized by a bulk scalar via the Goldberger-Wise mechanism, can give rise to a non-standard cosmology above a critical temperature $T_c \lesssim \mathcal{O}(\rm{TeV})$. It has been shown using the AdS/CFT correspondence that at high temperatures, the AdS-Schwarzchild geometry with an event horizon replacing the IR-brane is energetically favorable, and there is a first order phase transition to the RS geometry at $T_c$ as discussed in Ref. \cite{creminelli2002holography}. This appears to bring into question the viability of the thermally produced WIMP scenario in the RS background. However, once the phase transition into RS geometry has occurred, tunneling back into the AdS-Schwarzchild phase requires superheating at a temperature much larger than $T_c$ \cite{konstandin2011cosmological}. Thus when reheating occurs due to bubble collisions in the phase transition, we need not consider a constraint on the reheating temperature from $T_c$, and we can recover a standard cosmological evolution with a thermal relic WIMP. 

We also note that we neglect radion exchange in our calculation. Generically the radion is lighter than the first KK graviton, though its mass is not tied to the geometry in the same way as the KK graviton masses are, and it is a much more narrow resonance. This is due to both the radion being lighter as well as the radion's coupling to matter being weaker than that of the graviton by a factor of $1/\sqrt{3}$ in the notation of Ref. \cite{DHR}. As a result of the narrowness of the radion and its decreased coupling to matter, the cross-sections for radion exchange are greatly suppressed away from the radion resonance. Resonant $s$-channel exchange of the radion could potentially saturate the dark matter relic density, but only if the dark matter mass was almost exactly half of the radion mass. Taking $m_{radion} / m_{G1} \sim 2/3$ and comparing the cross-section for DM annihilating into an SM final state from $s$-channel exchange of an on-resonance radion to an on-resonance graviton, we find $\sigma_{radion}/\sigma_{graviton} \lesssim 1\%$. The Doppler broadening due to the thermal average serves to further suppress the narrow radion resonance. Since the radion may be lighter than the dark matter, annihilation into two radions could also be a relevant channel, as noted in Ref. \cite{Lee:2013bua}. However, the $t$- and $u$-channel diagrams do not benefit from the resonant behavior of the $s$-channel considered earlier, and thus are suppressed. Similarly, DM annihilation into two radions via a resonant $s$-channel radion exchange is not allowed kinematically, and so is also suppressed. Finally, there exists a four point interaction between two radions and two dark matter particles, but this channel does not benefit from any resonant enhancement while still having the same number of powers of the coupling as the previous channels, and is correspondingly suppressed. We thus concentrate our efforts on the KK graviton states as mediators of dark matter interaction with the SM. 

In the thermal freeze-out scenario, dark matter is produced in thermal equilibrium with the hot, dense early universe, and then freezes out as 
the universe expands and cools \cite{thermal}. In such cases the evolution of the number density of dark matter $n_\chi$ is governed by 
the Boltzmann equation:
\begin{equation}
\frac{d n_\chi }{dt} = -3Hn_\chi - \left< \sigma v_{M\o l} \right> \left[ n_\chi^2 - (n_\chi^{eq})^2 \right] \,,
\end{equation}
where $n_\chi^{eq}$ is the number density of DM particles that would exist in thermal equilibrium, $\sigma = \sum_{f \in SM} \sigma (\chi \chi \rightarrow f)$ is the sum of all annihilation channels into SM final states, $H(T)$ is the Hubble expansion rate as function of temperature T, and $v_{M \o l}$ is the M\o ller velocity as defined in Ref. \cite{gondolo1991cosmic}. In the early universe, the DM 
is in thermal equilibrium and the first term in the equation governs the number density evolution. As the universe expands and cools, the interaction rate $\Gamma_\chi \equiv n_\chi \left< \sigma v_{M \o l} \right>$ falls below $H(T)$ and the dark matter freezes out of thermal equilibrium and its number density in a co-moving frame remains (approximately) constant.

In this framework the thermal relic abundance of the dark matter is determined by the value of $\left< \sigma v_{M \o l} \right>$. 
Here we calculate this quantity analytically and do not use the velocity expansion which can be invalid in the neighborhood 
of any $s$-channel resonances; such contributions are an important region in our scenario as we will see in our discussion below. The thermal average is formally defined with each annihilating particle in its own thermal bath, and we can write $\left< \sigma v_{M \o l} \right>$ as
\begin{equation}
\left< \sigma v_{M \o l} \right> = \frac{\int \sigma v_{M \o l} e^{-E_1/T}e^{-E_2/T} d^3p_1 d^3p_2}{\int e^{-E_1/T}e^{-E_2/T} d^3p_1 d^3p_2} \,,
\end{equation}
where we have assumed $T \lesssim 3 m_\chi$, with $m_\chi$ being the dark matter mass, so that we may use Maxwell-Boltzmann statistics, following Ref. \cite{gondolo1991cosmic}. We have checked numerically that this is always an 
excellent approximation for the parameter ranges of interest to us here. In particular, the dimensionless ratio $x_f \equiv m_\chi / T_f$, where $T_f$ is the freeze-out temperature, typically takes on values of 20-30 \cite{nihei2002exact}. For simplicity in our analysis we set $x_f =$ 25. The above expression can be reduced to a one dimensional integral over the Mandelstam variable $s = (p_1 + p_2)^2$ where the $p_i$ are the 4-momenta of the initial DM states, giving
\begin{equation}
\left< \sigma v_{M\o l} \right> = \frac{1}{8 m_\chi^4 T K_2^2(m_\chi/T)} \int_{4 m_\chi^2}^{\infty} (s-4m_\chi^2) \sqrt{s} \sigma K_1\left(\frac{\sqrt{s}}{T}\right) ds \,.
\end{equation}
Here the annihilation cross section $\sigma$ is a function only of the masses of the various particles involved and $s$. We take  advantage of the frame-independence of this calculation and compute cross-sections in the center of mass frame. Taking $s = 4 m_\chi^2 \gamma^2$, where $\gamma$ is the familiar Lorentz factor $ \gamma = 1/\sqrt{1-\beta^2}$, with $\beta$ being 
the CoM DM velocity, with $T=T_f$, we can simplify the above expression to the form
\begin{equation}
\label{gamint}
\left< \sigma v_{M\o l} \right> = \frac{8 x_f}{K_2^2(x_f)} \int_{1}^{\infty} \sigma \gamma^2 (\gamma^2 - 1) K_1(2 x_f \gamma) d\gamma \,.
\end{equation}
All that remains to be determined are the individual  annihilation cross sections $\sigma$. 

These annihilation cross sections are calculated using the KK graviton Feynman rules found in Refs. \cite{han1999kaluza, DHR}, 
employing modifications to account for the factors of $\delta_n$ defined in Eq. \ref{del} which appear in front of the 5D 
Yang-Mills kinetic terms. Similarly, for the various chiral fermions, we modify their couplings by factors of $c_{L,R}$ to 
account for their localization (brane vs. bulk) that can differ for SM doublet and singlet fields of the same flavor and  
which are accompanied by the projection operators $P_{R,L} = \frac{1}{2}(1 \pm \gamma_5)$. In particular, we write the Feynman rule for the three point interaction between the $n^{th}$ KK graviton and two on-shell vector bosons as
\begin{equation}
\frac{-i}{\sqrt{2} \Lambda_\pi} \delta^{ab} \lambda_n \left[ m_V^2 C_{\mu \nu, \rho \sigma} + \delta_n \left( k_1 \cdot k_2 
~C_{\mu \nu, \rho \sigma} + D_{\mu \nu, \rho \sigma}(k_1, k_2) \right) \right] \,,
\end{equation}
where $C_{\mu \nu, \rho \sigma}$ and $D_{\mu \nu, \rho \sigma}(k_1, k_2)$ are defined in the Appendix of Ref. 
\cite{han1999kaluza}, $k_1$ and $k_2$ are the 4-momenta of the respective vector bosons, $m_V$ is the mass of the vector boson, 
$\lambda_n$ is the coupling strength to the graviton defined above in Eqn. \ref{lambdaeq}, and $\delta^{ab}$ is the Kroenecker  
delta function of the gauge indices $a$ and $b$ of the vector bosons. Note that the factor of $\delta_n$ only modifies the terms in the Feynman rule
arising from the Yang-Mills kinetic term, and it does not modify the term that is proportional to the boson mass, which is generated via spontaneous symmetry breaking on the IR-brane.

Next we consider the coupling of the KK graviton states to the chiral fermions. For the right-handed singlet $f_R$, we write the Feynman rule for the {\it kinetic part} of the on-shell three point interaction as  
\begin{equation}
\frac{-i}{2 \sqrt{2} \Lambda_\pi} \delta^{ab} c_{R,n} \lambda_n P_R  \left[ \gamma_\mu (k_{1 \nu} + k_{2 \nu}) + \gamma_\nu (k_{1 \mu} + k_{2 \mu}) \right] \,,
\end{equation}
where $k_1$ (-$k_2$) is the 4-momentum of the incoming (anti)fermion, 
$\delta^{ab}$ is the Kroenecker delta function for the fermion flavor indices, and $c_{R,n}$ is the modification to the 
coupling strength $\lambda_n$ arising from the possible localization of the right-handed fermion in the 5D bulk. In the case of a 
fermion residing on the TeV brane, $c_{R,n} = 1$. The Feynman rule for the left-handed fermions is simply obtained by replacing 
$P_R$ and $c_{R,n}$ with $P_L$ and $c_{L,n}$. It is important to remember that the right- and left-handed chiral fermions may 
be located at different points in the 5D bulk, so that in general $c_{R,n} \neq c_{L,n}$. In addition to these kinetic pieces, 
there is also a coupling term on the IR-brane which corresponds to the fermion mass generated by spontaneous symmetry 
breaking via the Higgs vev and is simply proportional to $m_f\lambda_n \eta_{\mu\nu}$, linking the two 
chiral fermion fields.  

Finally we consider the three point interaction between two incoming scalars and the $n^{th}$ KK graviton. We write the Feynman rule as 
\begin{equation}
\frac{-i}{\sqrt{2} \Lambda_\pi} \delta^{ab} \lambda_n (m_s^2 \eta_{\mu \nu} - C_{\mu \nu, \rho \sigma} k_1^\rho k_2^\sigma) \,,
\end{equation}
where $m_s$ is the mass of the scalar, $k_1$ and $k_2$ are the 4-momenta of the incoming scalars, and $\lambda_n$, $\eta_{\mu \nu}$ and $C_{\mu \nu, \rho \sigma}$ are previously defined. Note that since we only consider scalars being confined to the TeV brane, the Feynman rule in Ref. \cite{han1999kaluza} is only modified by a factor of $\lambda_n$. 

\begin{figure}
\begin{center}
\includegraphics[scale=0.8]{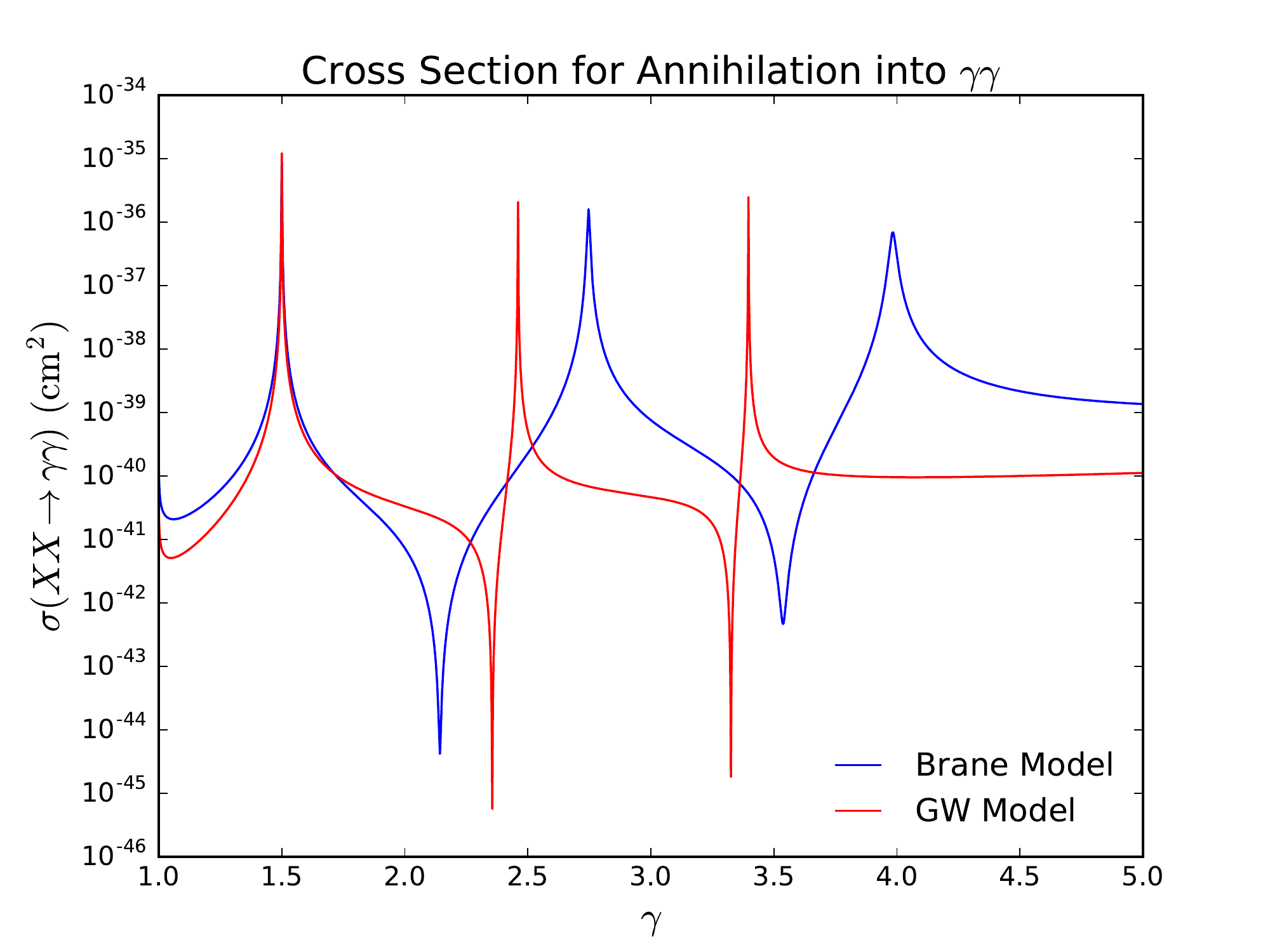}
\end{center}
\caption{The cross section for spin-1 dark matter annihilating through $s$-channel exchange of the first 3 KK gravitons as a function of the Lorentz factor $\gamma = \sqrt{s}/2 m_X$ for the GW (red) and Brane (blue) models, taking $m_X$ = 1000 GeV and $m_{G_1}$ = 3000 GeV. Note in particular that the weaker couplings of the higher KK gravitons to matter in the GW model push the maximal interference conditions close to the $s$-channel resonance peaks.} 
\label{factorfig}
\end{figure}

Since all KK graviton states can mediate a given annihilation process, it is important to consider the effects of interference between the KK modes. The interference effects between the $i^{th}$ and $j^{th}$ KK gravitons appear in the cross sections via the factor

\begin{equation}
\label{pij}
P_{ij} = \frac{(s-m_{G_i}^2)(s-m_{G_j}^2) + m_{G_i} \Gamma_i m_{G_j} \Gamma_j}{[ (s-m_{G_i}^2)^2 + m_{G_i}^2 \Gamma_i^2 ] [  (s-m_{G_j}^2)^2 + m_{G_j}^2 \Gamma_j^2 ]} \,,
\end{equation}
where $m_{G_i}$ and $\Gamma_i$ are the mass and width of the $i^{th}$ KK graviton. The cross sections contain the sum of these interference factors, $ \sum_{i,j} P_{ij}$. When the KK gravitons have different coupling strengths, factors of $\lambda_i^2 \lambda_j^2$ (along with factors of $c_{R/L,i} c_{R/L,j}$ for the case of chiral fermions or $\delta_i \delta_j$ for vector bosons) appear as well. The impact of the interference factors can be seen in Fig. \ref{factorfig} which shows the annihilation into two photons. In the case of the Brane model we have $\lambda_i = 1$ for all $i$, and the interference minima are roughly evenly spaced between KK graviton resonances. When the KK gravitons couple to matter with different strengths, the interference minimum moves closer to the KK graviton resonance which couples more weakly to matter. In the case of the GW model we have $\lambda_1 > \lambda_2 > \lambda_3$ etc., so the interference minima are located near the heavier KK graviton resonance peaks.

The annihilation rates into the various SM final states contain velocity 
suppression factors resulting from the choice of spin of the DM particle. As an example, consider the cross sections for scalar, fermionic, 
and vector DM annihilating into a pair of Higgs bosons localized on the TeV brane (as given in Eqs. \ref{scalsec}, 
\ref{fermsec}, and \ref{vecsec} of the Appendix, respectively). For the case of scalar DM, $\sigma v$ is 
proportional to a factor of $\beta_S^4$, which corresponds to $\sim v^4$ in the center of mass frame. This represents so-called ``d-wave'' suppression of the annihilation since for spinless initial 
state DM we need to be in the $L=2$ angular momentum state to match the $J=2$ spin of the intermediate graviton KK 
state. Similarly, $\sigma v$ for fermionic DM annihilation is proportional to $\beta_F^2$, 
corresponding to $\sim v^2$ in the CoM, and is thus ``p-wave'' suppressed since the fermionic DM must be 
in the $L=1$ state to match with the KK graviton spin. In the case of vector DM, however, $\sigma v$ is not suppressed by 
any factors of $\beta_V$, as in this case annihilation can take place in the $L=0$ state or ``s-wave'' 
annihilation. It is important to note that these velocity-dependent suppressions are completely generic - all 
amplitudes for spin-0(1/2, 1) particles coupling to the graviton will experience an overall scaling of $\beta_S^2$ 
($\beta_F^1$, $\beta_V^0$). Similar velocity suppressions are also present for the various SM final states, corresponding to their 
associated spins. E.g., in the case of annihilation to Higgs boson pairs, the cross section is always suppressed by a factor of 
$\beta_h^5$ (which includes the phase space correction) independent of the initial DM spin. Furthermore, in the simplest case where $SS\to hh$, so that both initial and final states must be in the d-wave, we not only see the appropriate velocity suppressions but also the factor of $(1-3z^2)^2$ 
(where $z=\cos \theta$ being the CoM scattering angle). This corresponds to the square of the second order Legendre polynomial  
as expected as the $L=2$ orbital angular momentum wavefunction.  This type of velocity scaling behavior is envisioned on general grounds based on the Breit-Wigner approximation.  

\begin{figure}
\begin{tabular}{l l}
\hspace{-1.65cm}\includegraphics[scale=0.45]{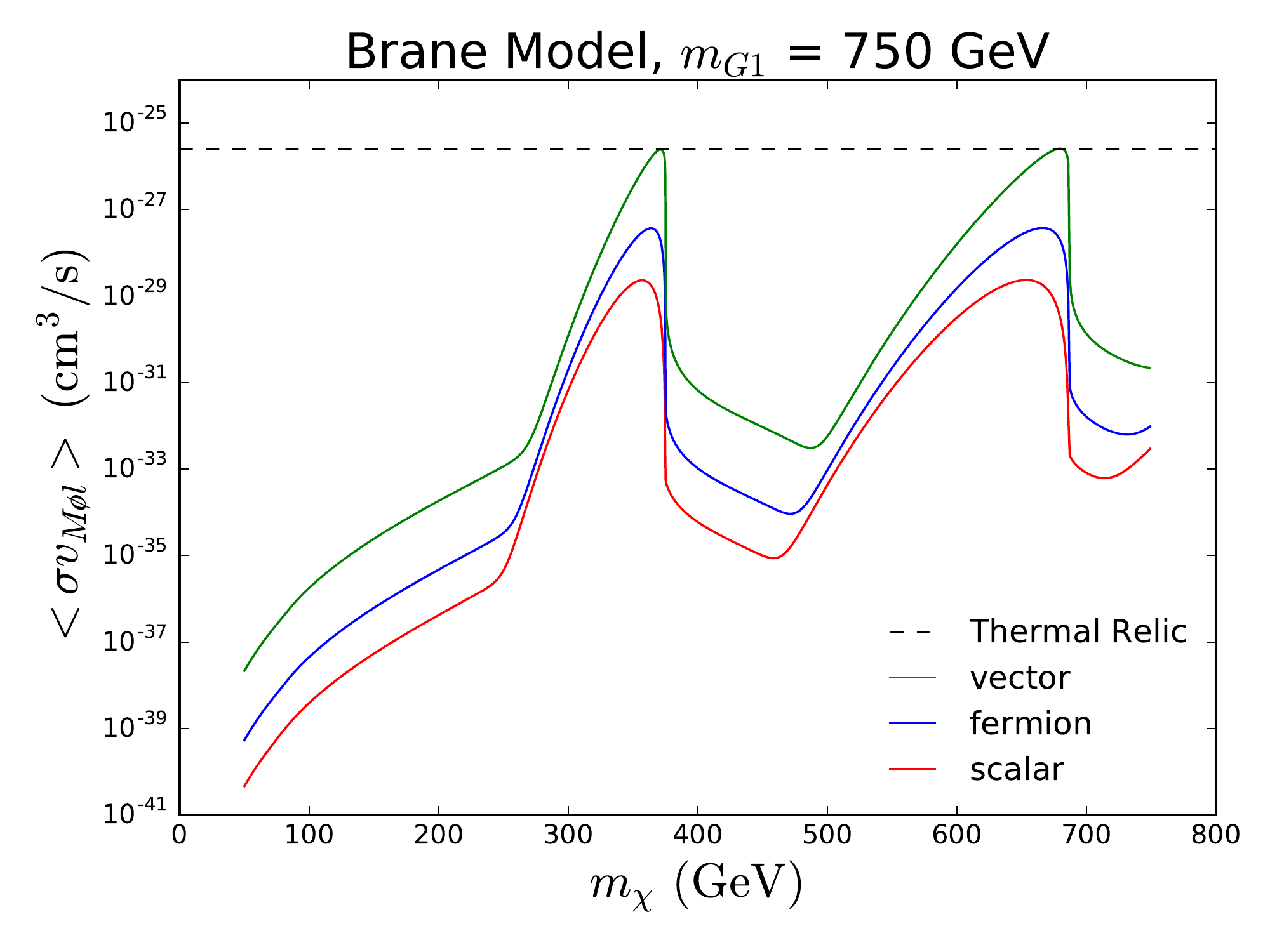}&\includegraphics[scale=0.45]{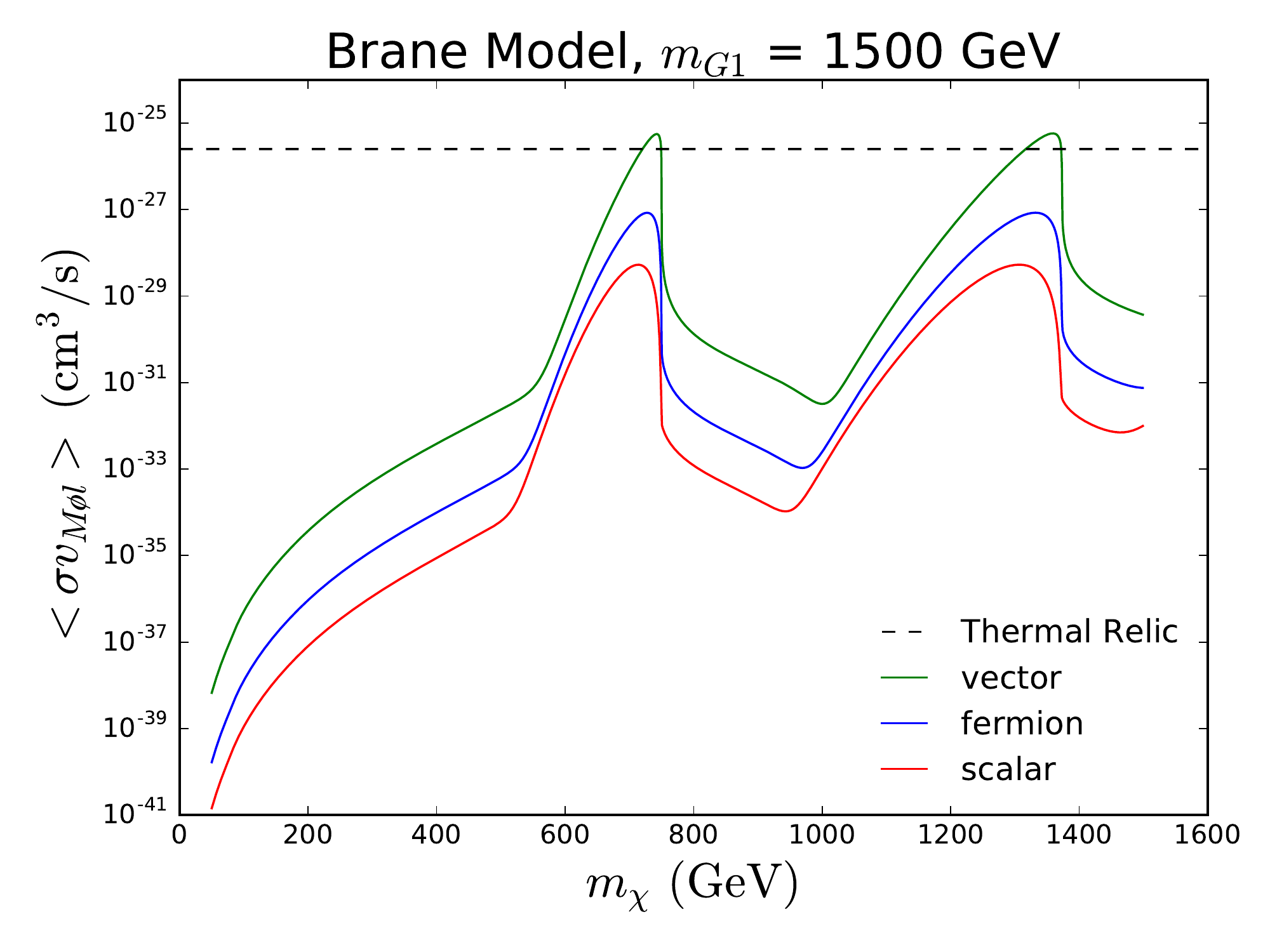}
\end{tabular}
\begin{center}
\vspace{-0.5cm}
\includegraphics[scale=.45]{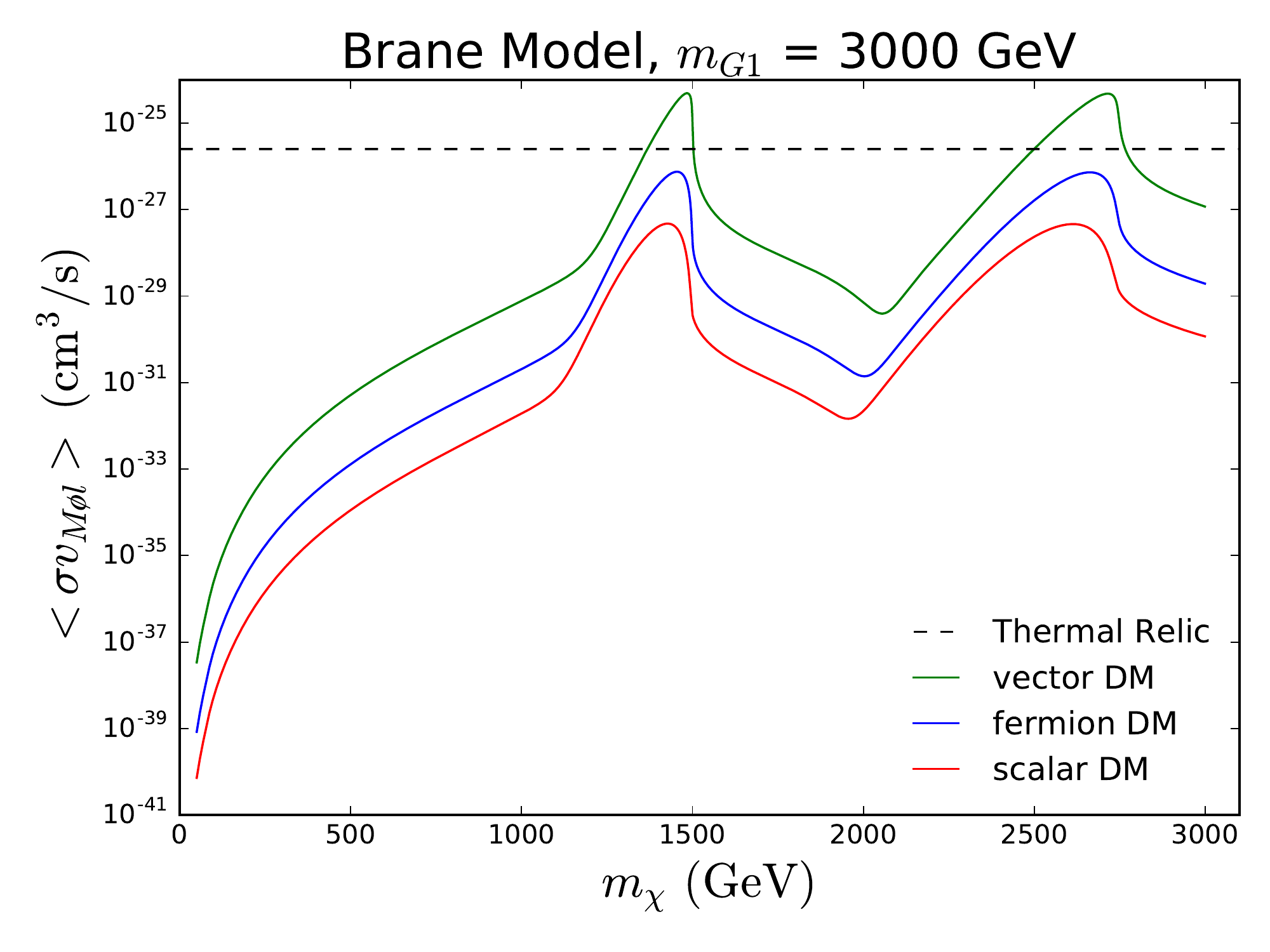}
\end{center}
\caption{The thermally averaged cross section times M\o ller velocity for the benchmark models with all SM fields residing on the IR-brane as a function of the DM mass. The three benchmarks correspond to a lowest KK graviton mass of 
750 GeV (top left), 1500 GeV (top right), and 3000 GeV (bottom). The value of $\left< \sigma v \right>$ associated with the 
observed relic abundance of DM is shown by the black dashed line.}
\label{branefig}
\end{figure}

\begin{figure}
\begin{tabular}{l l}
\hspace{-1.65cm}\includegraphics[scale=0.45]{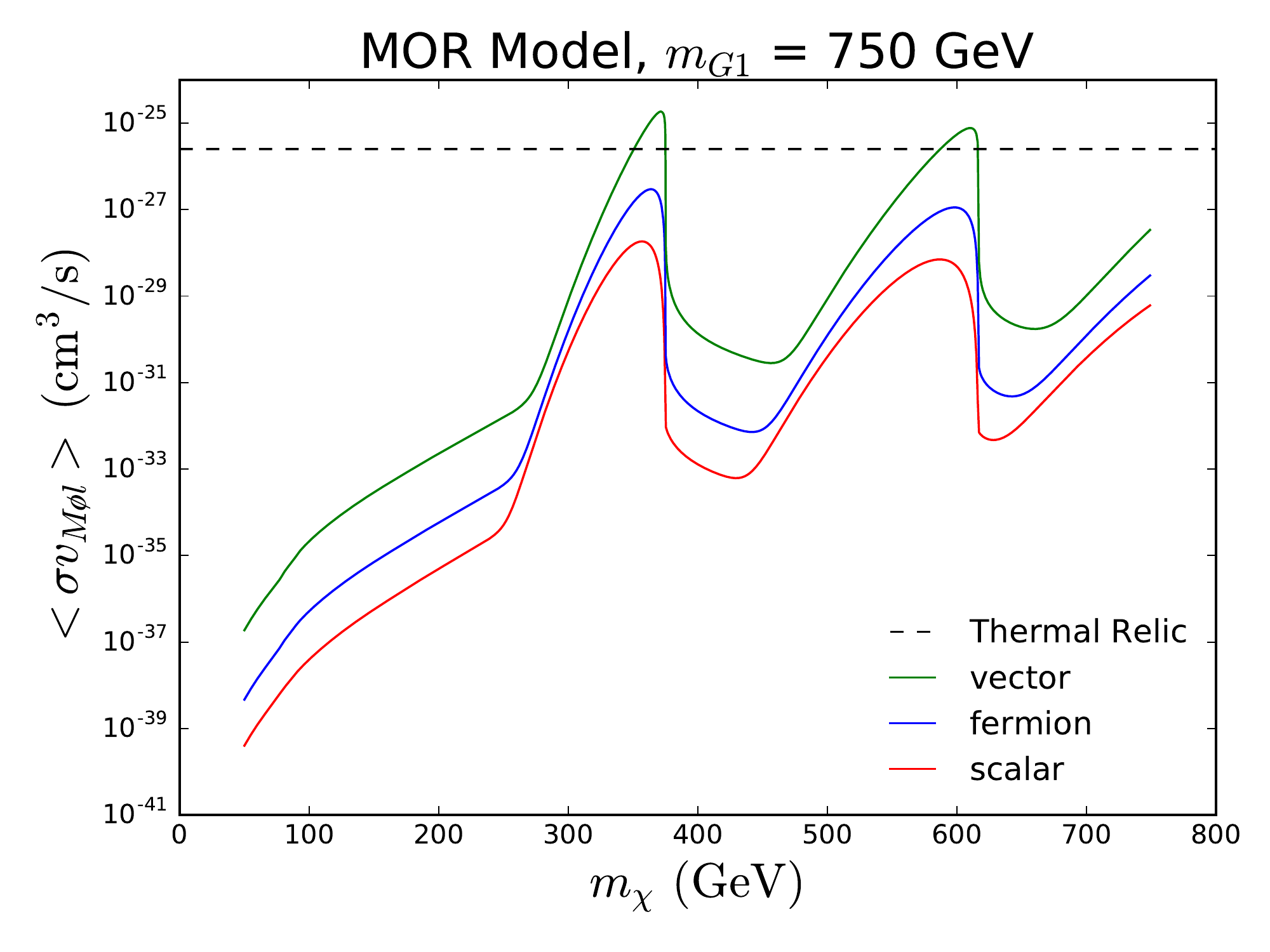}&\includegraphics[scale=0.45]{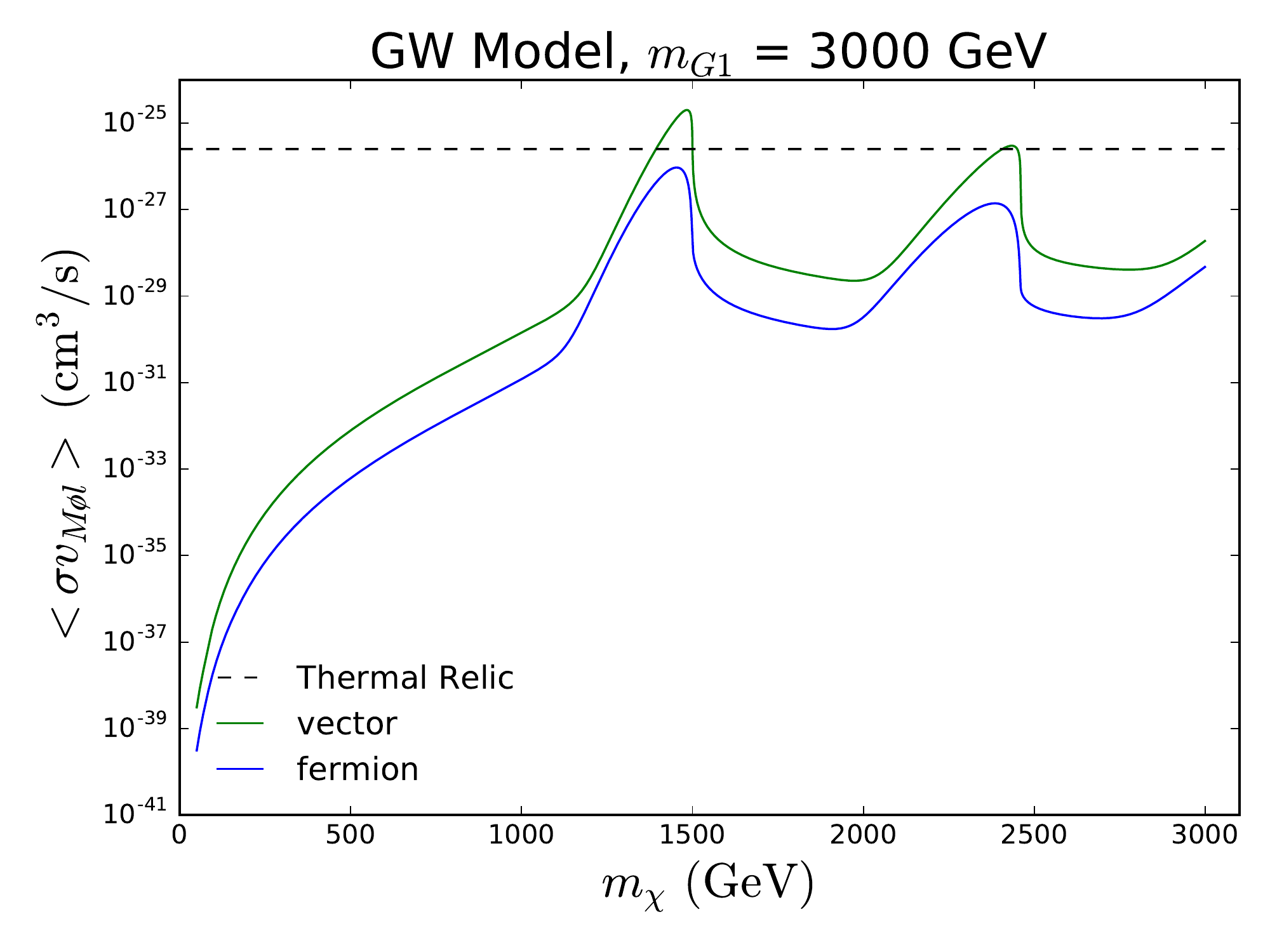}
\end{tabular}
\caption{The thermally averaged cross section times M\o ller velocity for the MOR (left) and GW (right) benchmark models. 
The value of $\left< \sigma v \right>$ associated with the observed thermal relic abundance of DM value is shown by the black 
dashed line.}
\label{gwmorfig}
\end{figure}

Expressions for the cross sections for DM annihilation into the various particles of the SM are given in Appendix \ref{xsecapp}. Given each of 
these individual contributions we can now calculate the inclusive $\left< \sigma v_{M \o l} \right>$ for all of the 
various benchmark models described in Section \ref{section:Model}. We calculate the integral in Eq. \ref{gamint} 
numerically, using the VEGAS algorithm for adaptive Monte Carlo integration \cite{lepage1978new}. The results for each 
benchmark model are shown in Figs.~\ref{branefig} and ~\ref{gwmorfig}. 

We first examine the benchmark models with all the SM particles living on the IR-brane. The velocity suppression 
factors for the scalar and fermionic DM cases discussed above prevent these possibilities from saturating the observed relic abundance and creates the hierarchy of $\left< \sigma v_{M \o l} \right>$ values 
seen in Fig. \ref{branefig}. Note that as the mass of the lightest graviton KK increases, the cross sections near 
KK resonances increase due to the non-renormalizable graviton coupling. Thus while we see that the scalar and fermion 
DM scenarios never reach the desired value of the relic density for any of our benchmarks, we note that this preferred value may be reached in the case of fermionic DM if the lightest KK graviton mass is increased to 4 TeV.
However, for the specific IR-localized SM benchmarks that we have chosen the DM can only be spin-1, with mass near $m_{G_n}/2$, in order to avoid over-closing the universe. 
We further observe that the second KK excitation peak cross section is quite close to that of the first KK state; we will return to 
this point below. 
It is also interesting to note that the benchmark scenarios that saturate the observed relic abundance only do so 
near the $s$-channel KK resonances where $m_{DM} \simeq m_{G_n}/2$. The broadness of the peaks in 
$\left< \sigma v_{M \o l} \right>$ arises from the Doppler broadening which occurs as part of the thermal averaging process - 
the widths of the KK gravitons themselves, as calculated in Appendix \ref{appwidth}, are actually quite narrow. As an 
example, when $ m_{G_1} = 750$ GeV, we find that $\Gamma_{G_1} \simeq 2$ MeV, but the thermal averaging makes the 
resonance accessible to a significant amount of the DM thermal distribution for lower masses, thus broadening the first peaks 
in Fig. \ref{branefig}. The steep drop off in the annihilation rate at $m_{DM} = m_{G_n}/2$ is a  
result of the Breit-Wigner peak in the cross section falling below the accessible center of mass energy, 
\ie, falling below $2 m_{DM}$. 

The MOR and GW benchmark models exhibit very similar behavior to the brane models discussed above, though the 
differing coupling strengths of the KK gravitons to matter create some important differences with respect to the thermal 
peaks. As in the IR-brane localized benchmark cases, we again see that only the possibility of spin-1 DM can reach the 
required thermal relic cross section although, increasing the lowest graviton KK mass may open up other possibilities. 
Here, in addition to the interference between KK modes being suppressed in the thermal averaging process as explained above, 
the relative peak heights within a single benchmark model are modified by the KK-number dependent, non-universal couplings. 
In particular, the peak heights of higher KK modes are reduced due to their smaller couplings to SM matter. This is most 
notable in the GW model shown in the right-hand panel of Fig. \ref{gwmorfig}, where the second peak just rises to the observed thermal 
relic abundance cross section despite the first peak exceeding it by a substantial margin. 

The interference effects between successive KK graviton states are also important to understand the shape of the various   
annihilation curves. In particular, the interference condition between the first two KK graviton modes is 
clearly visible in the Brane models as a strong dip in  $\left< \sigma v_{M \o l} \right>$ near $m_{DM} \sim 0.75 m_{G_2}$. This effect is most visible in the case of $m_{G_1} = 3000$  GeV in Fig. \ref{branefig} bottom. This feature is much less prominent in the MOR and GW models, shown in Fig. \ref{gwmorfig}, due to the non-universal couplings. The Boltzmann factors in the thermal average suppress contributions of the cross section at high $s$, so only the low $s$ portion of the cross section contributes to the integral for a given DM mass. As a result, when there is large spacing (in $s$) between the interference minima and resonance maxima, it is possible to pick up a single minimum or maximum, which is what occurs in the Brane model with $m_{G_1} = 3000$ GeV in Fig. \ref{branefig} bottom. When the maxima and minima are pushed near one another due to KK gravitons coupling to matter with different strengths, the thermal average smears out the interference minimum with the resonance. This so-called Doppler broadening is responsible for the absence of dips in $\left< \sigma v_{M \o l} \right>$ in the MOR and GW models in Fig. \ref{gwmorfig}.

As already noted, all of these models are capable of saturating the observed relic abundance of dark matter in the case of 
spin-1 DM, as evidenced by the intersection of the green curves with the line corresponding to the thermal relic abundance 
cross section. We emphasize, however, that this only occurs for DM masses in the neighborhood of the $s$-channel resonances of the KK 
gravitons. The velocity suppression of the p-wave and d-wave annihilations (corresponding to fermionic and scalar DM, 
respectively) prevents them from achieving a sufficiently large value of $\left< \sigma v_{M \o l} \right>$ in all of our 
benchmark models, thus leading to a predicted overabundance of DM in these scenarios. Again we emphasize, however, that for 
larger values of the lightest KK graviton mass $m_{G_1}$ beyond our benchmark values, the fermion DM possibility (at least) can become a viable 
scenario for thermal DM.

\subsection{Dark Matter Indirect Detection}

The properties of DM, as well as those of the KK gravitons in the models considered here, can in principle be constrained 
by searches for 
the various signatures that result from the DM annihilation process. Since DM annihilation proceeding through KK graviton exchange    
essentially leads to most, if not all, of the kinematically accessible SM particles with respectable branching 
fractions (BFs), there are numerous final states that can be employed in this endeavor. Here we will limit ourselves to the 
$\gamma \gamma, e^+e^-, \bar pp$ as well as the photon continuum (which follows from, \eg, the fragmentation process) final states 
as these potential signal channels cover most of the important graviton decay modes, and correspond to the experimental signatures. Interestingly, for the set of 
benchmark cases where all of the SM fields are confined to the IR-brane (and similarly for the GW scenario as well), 
apart from possible contributions that arise from decays to lighter gravitons that can occur for the higher KK excitations, the KK gravitons have the same 
BFs to the relevant SM fields, allowing for small phase space 
corrections. Thus, below the $G^3 \to 2G^1$ threshold and above top pair threshold, the ratio of, \eg, the DM annihilation 
branching rate into the diphoton final state is roughly a constant and is independent of the spin of the 
DM. This ratio can be identified with the value of $B(G^1\to 2\gamma) \simeq 0.042$ (in the case of IR-brane localization), 
as can be seen from a consideration of the Breit-Wigner approximation for the various $s$-channel annihilation cross sections 
employing running decay widths. 

In order to access the impact of present-day indirect detection searches on our model parameter space, we must re-examine 
the thermal relic calculation of $\left< \sigma v_{M \o l} \right>$ performed above.  
The reason for this is clear: instead of being semi-relativistic at freeze-out, the DM of interest to us at present times
is gravitationally bound in galaxies and is quite non-relativistic, \ie, $v \sim 10^{-3}$.  The present time DM velocity distribution 
roughly takes the form of a truncated Maxwellian
\begin{equation}
f(v) \sim ~v^2 ~e^{-3v^2/2v_0^2} ~\theta(v_{max}-v)\,,
\end{equation}
where $v_0$ and $v_{max}$ are both typically of $O(10^{-3})$, and $\theta(x)$ is the Heaviside theta function. (Here, for numerical purposes in our  calculations below we 
take $v_0(v_{max})=150(550)$ km/sec.{\cite{Kafle:2014xfa}}, although this specific choice of values turns out not to 
be very important as discussed below.) However, in practical applications, one finds that since the typical value of the 
velocity is sufficiently small, the approximation $v \to 0$ is an excellent one in performing the necessary cross 
section calculations below for the case of spin-1 DM.{\footnote {We have explicitly checked that employing either 
$v=0$ or the distribution above with finite DM velocities, produces the same result in this case at 
the $\sim 1\%$ level.}}  Due 
to the large $\sim v^2(v^4)$ suppression factors experienced in the spin-1/2(0) DM scenarios, the annihilation 
cross sections at present times for these cases  
are highly suppressed by many orders of magnitude and do not reach the required thermal cross sections even at freeze-out. 
The resulting annihilation rates in these two cases for all of the various signal 
final states are found to be very far below those probed by the current indirect search experiments (as we will see below) and so 
are not constrained by them, or by any foreseeable improvements in these searches in the near future. Thus we need not consider 
these two spin cases any further here. Let us now turn to a discussion of the indirect detection final 
states relevant for experiment, and focus our attention on the spin-1 DM case. As we will discover, independent of the search channel, these constraints 
are extremely weak at best. 

The $\gamma \gamma$ `photon-line' search mode is potentially more important for the KK graviton-mediated DM 
annihilation process 
than in the case of, \eg, neutralinos in Supersymmetry where such a final state is only achievable via 1-loop graphs and correspondingly 
has a highly suppressed diphoton branching fraction. As noted above, by contrast, the diphoton branching fraction for KK gravitons lie in the range of 
$\sim 4-8$\%, depending upon how the various SM fields are localized. The existing limits on this mode arise from both the  
Fermi-LAT{\cite {Ackermann:2015lka}} and HESS{\cite {Abramowski:2013ax,Abdalla:2016olq} Galactic Center (GC) data sets, with 
the actual constraints being strongly dependent upon whether a cored or cusped DM profile is assumed to be 
applicable at the GC. Future measurements by CTA {\cite {Lefranc:2016fgn}} are expected to significantly improve 
the existing constraints over a reasonable DM mass range. In Figs.~\ref{cusp} and ~\ref{core} we compare these 
observational constraints for this final state with the present day ($z=0$) DM annihilation rates, and, for 
contrast, the rate at thermal freeze-out for our five benchmark 
models, assuming both cusped and cored DM profiles. These sets of distributions have several features in common: As already 
noted above, the Doppler broadening in the region of the graviton resonance peaks produce a rather wide maximum in the 
annihilation cross section during freeze-out. Today, DM is so slowly moving that $v \simeq 0$ is a good approximation 
in which case the width of the resonance peak is now set by that of the graviton KK resonances themselves, which 
are quite narrow for all 5 of our benchmarks.  In the case of DM with a cusped profile, it is clear that these constraints 
can potentially be important in the very narrow mass range around the resonances for all 5 benchmark models, however, most of the 
parameter space allowing for the observed relic density is not impacted by these constraints. On the other hand, employing  
the cored DM profile yields {\it no} constraints from these observations. The reason for this is 
the predicted  photon flux from the GC is proportional to the square of the DM density integrated over the region of interest 
on the sky, and so is much reduced in the case of a cored profile. However, for either choice of DM profile, the impact of 
these searches on our parameter space is at most quite minimal and only very close to a KK resonance.  We note that improved sensitivity
is expected in the future from searches at CTA.

\begin{figure}
\vspace{-0.7cm}
\begin{tabular}{l l}
\hspace{-1.65cm}\includegraphics[scale=0.45]{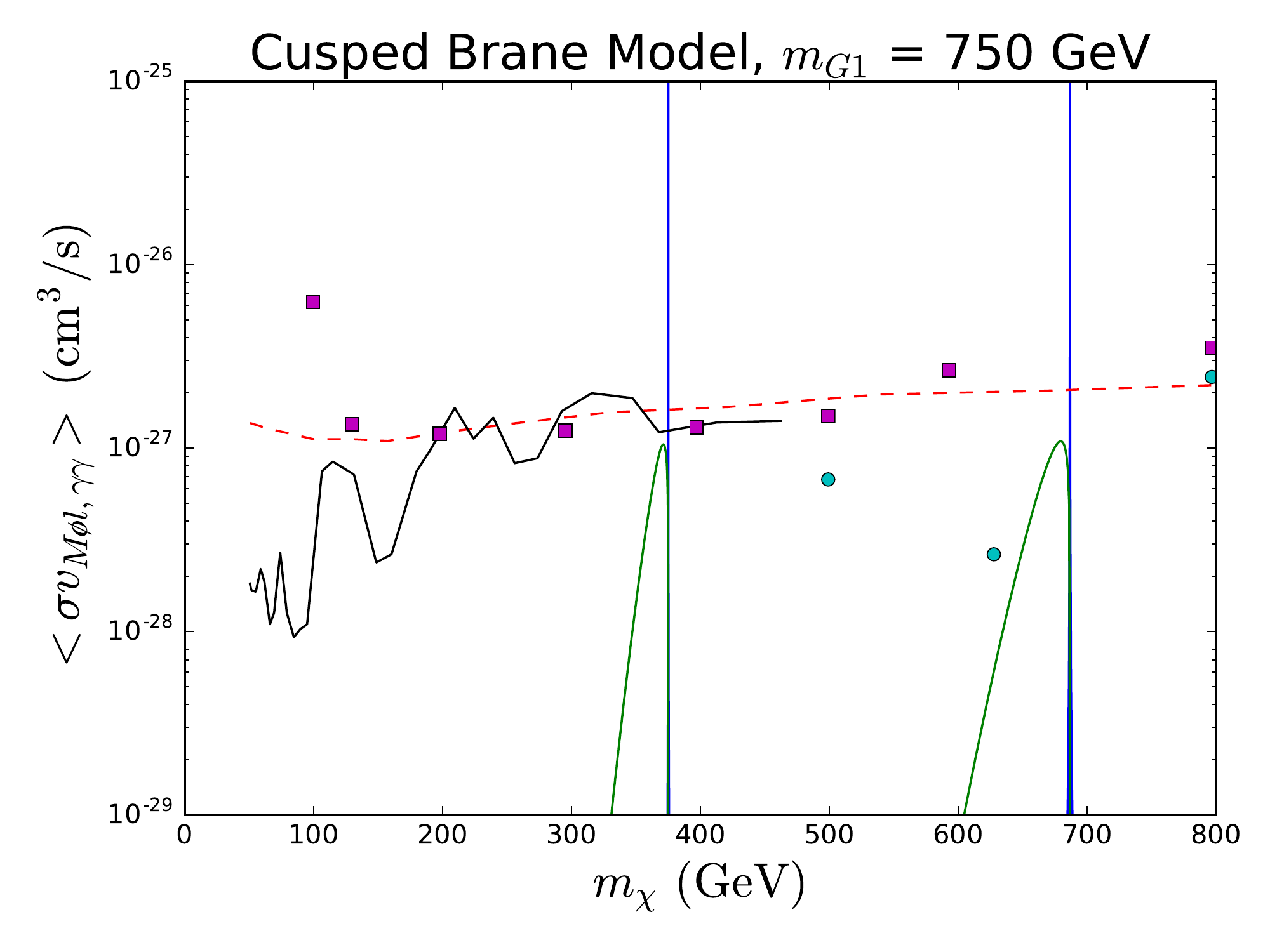}&\includegraphics[scale=0.45]{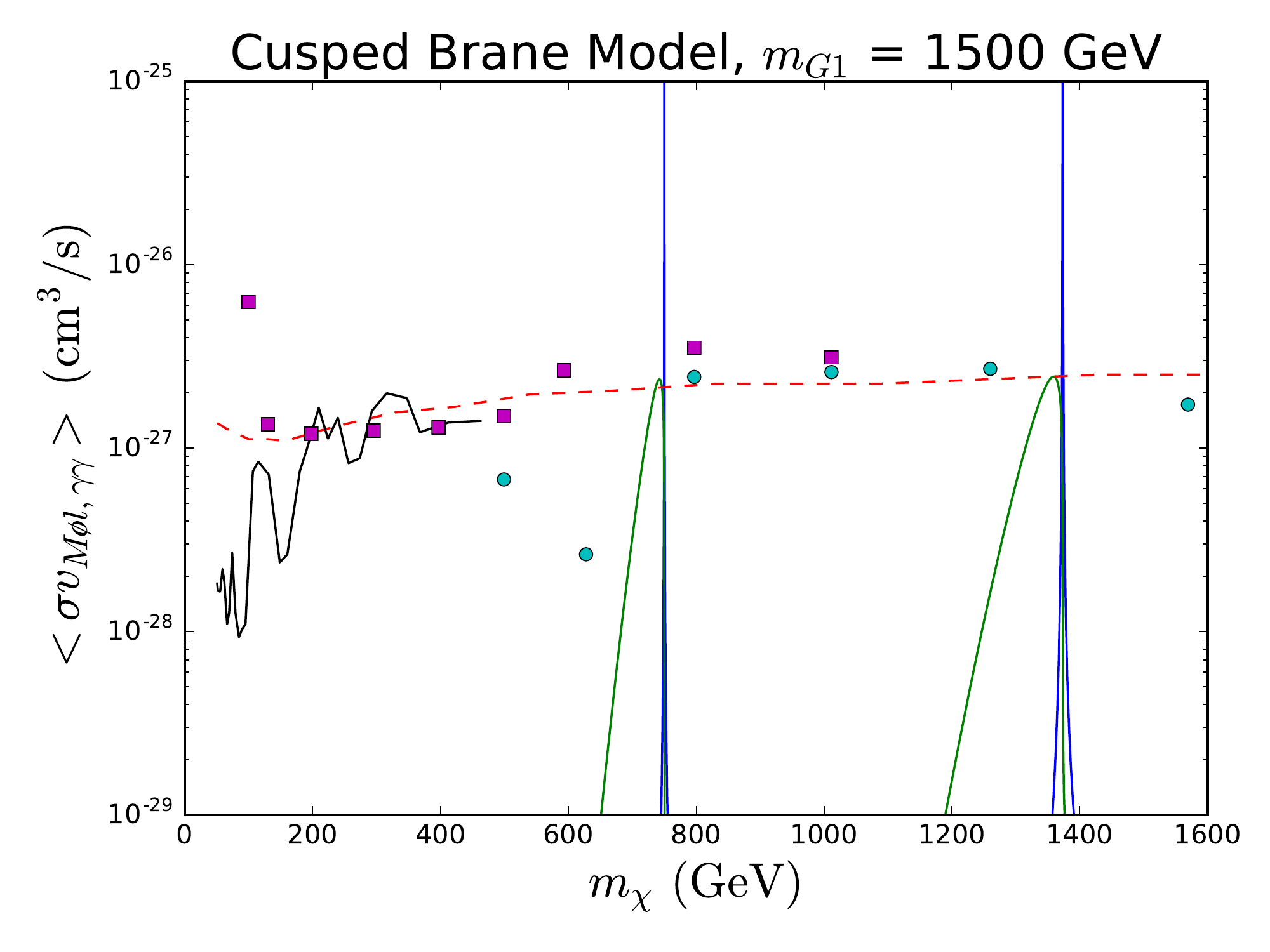} \\
\hspace{-1.65cm}\includegraphics[scale=0.45]{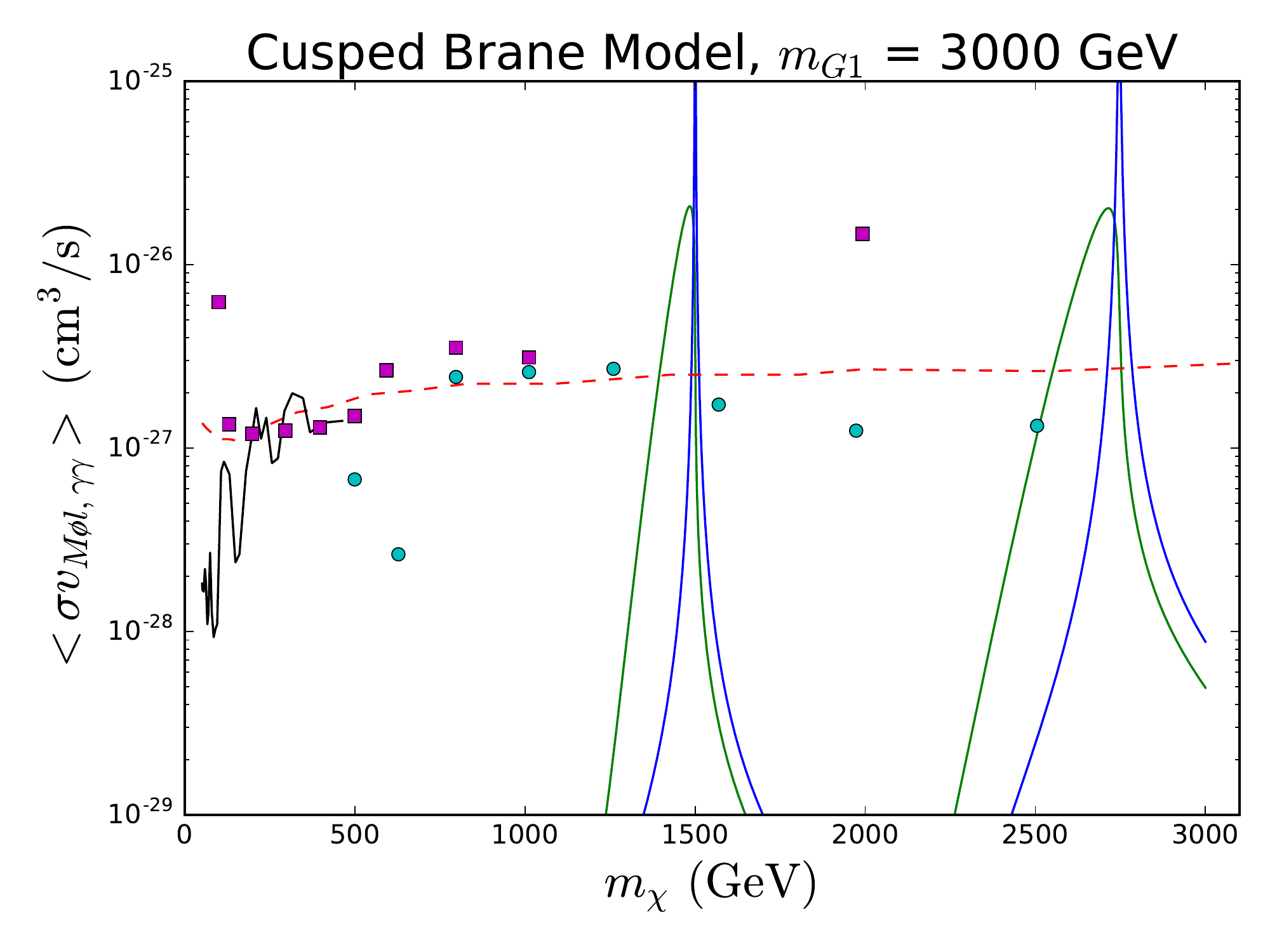}&\includegraphics[scale=0.45]{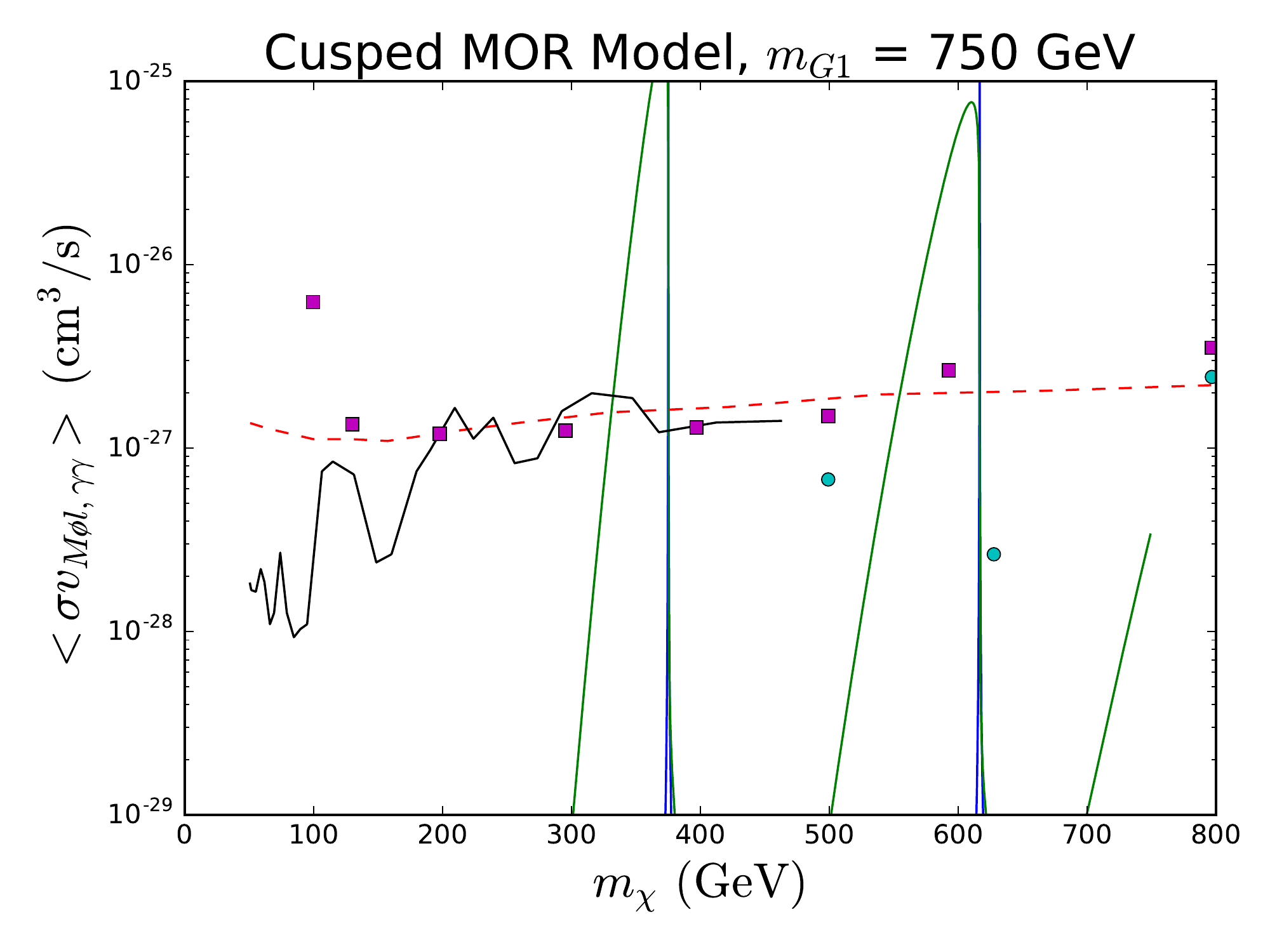}
\end{tabular}
\begin{center}
\vspace{-0.5cm}
\includegraphics[scale=.45]{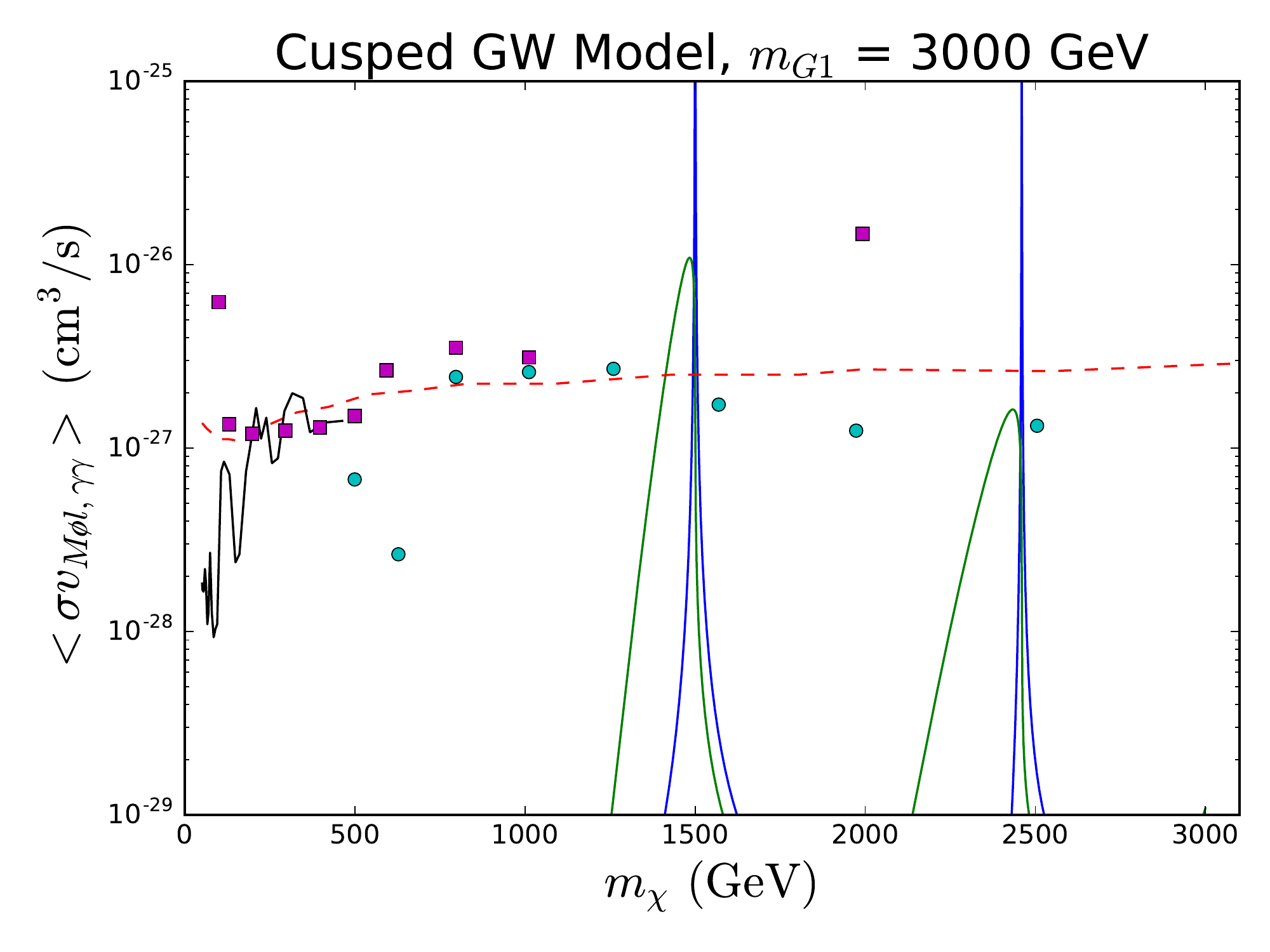}
\end{center}
\vspace{-0.5cm}
\caption{Assuming a cusped profile, the thermally averaged dark matter annihilation cross section into two photons at $z=0$
(at thermal freeze-out) for spin-1 dark matter as a function of the DM mass is given by the blue (green) curve for the 5 benchmark models as labeled.
The photon line search limits from FERMI (solid black), HESS I (cyan circles), and 
HESS II (magenta squares) are also shown, as well as the projected 
sensitivity of the Cerenkov Telescope Array (red-dashed line). }
\label{cusp}
\end{figure}

\begin{figure}
\vspace{-0.7cm}
\begin{tabular}{l l}
\hspace{-1.65cm}\includegraphics[scale=0.45]{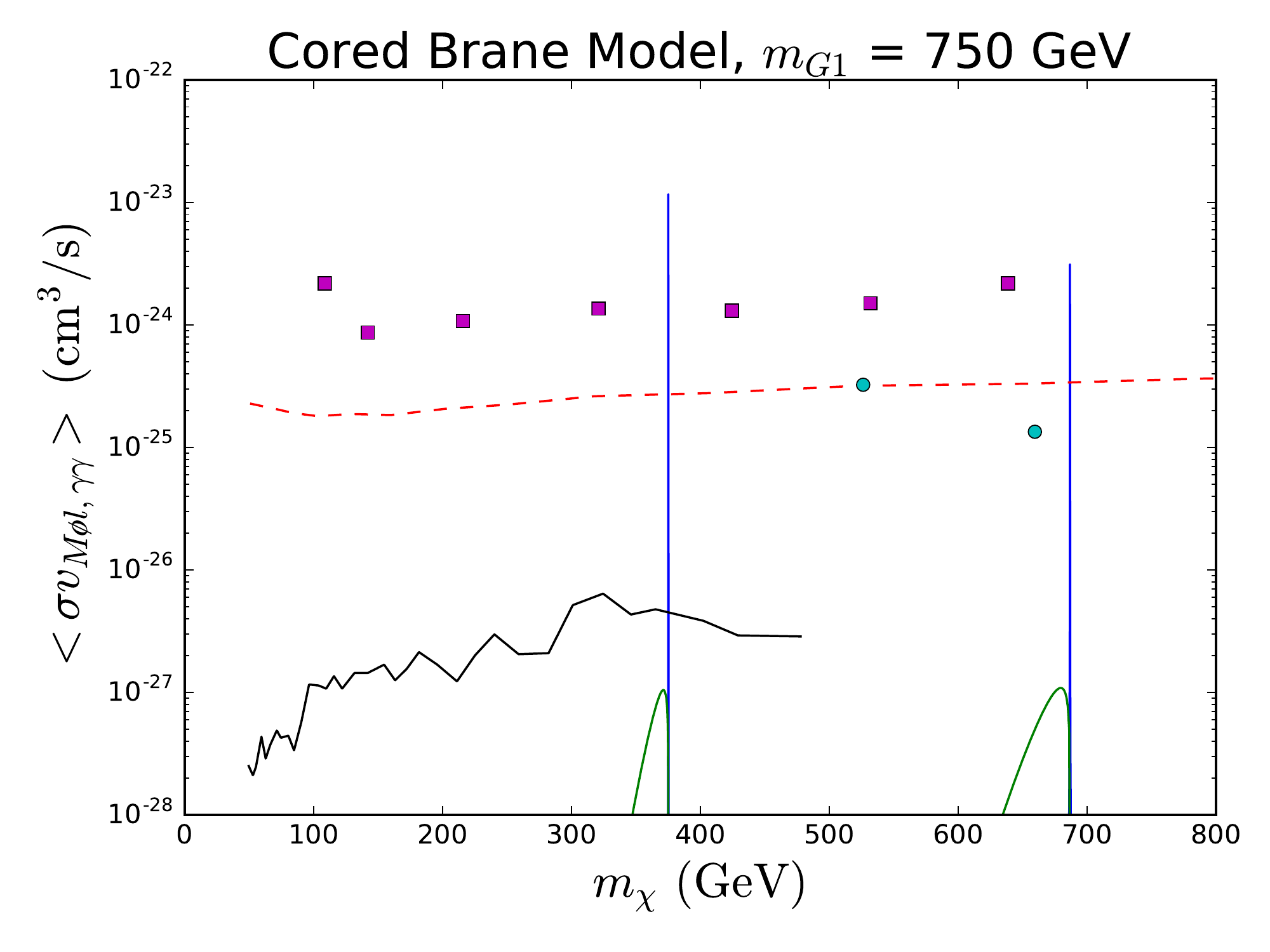}&\includegraphics[scale=0.45]{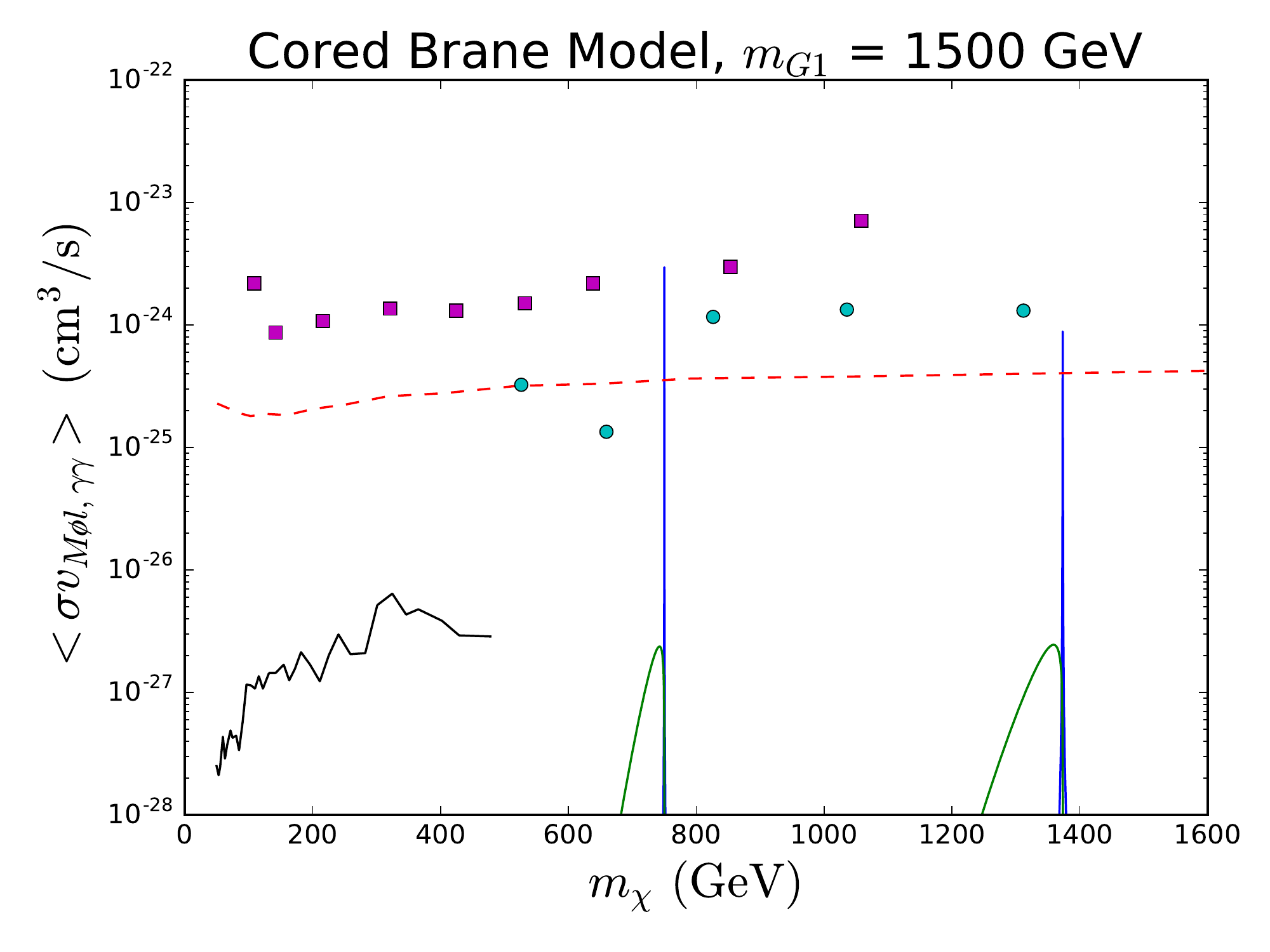} \\
\hspace{-1.65cm}\includegraphics[scale=0.45]{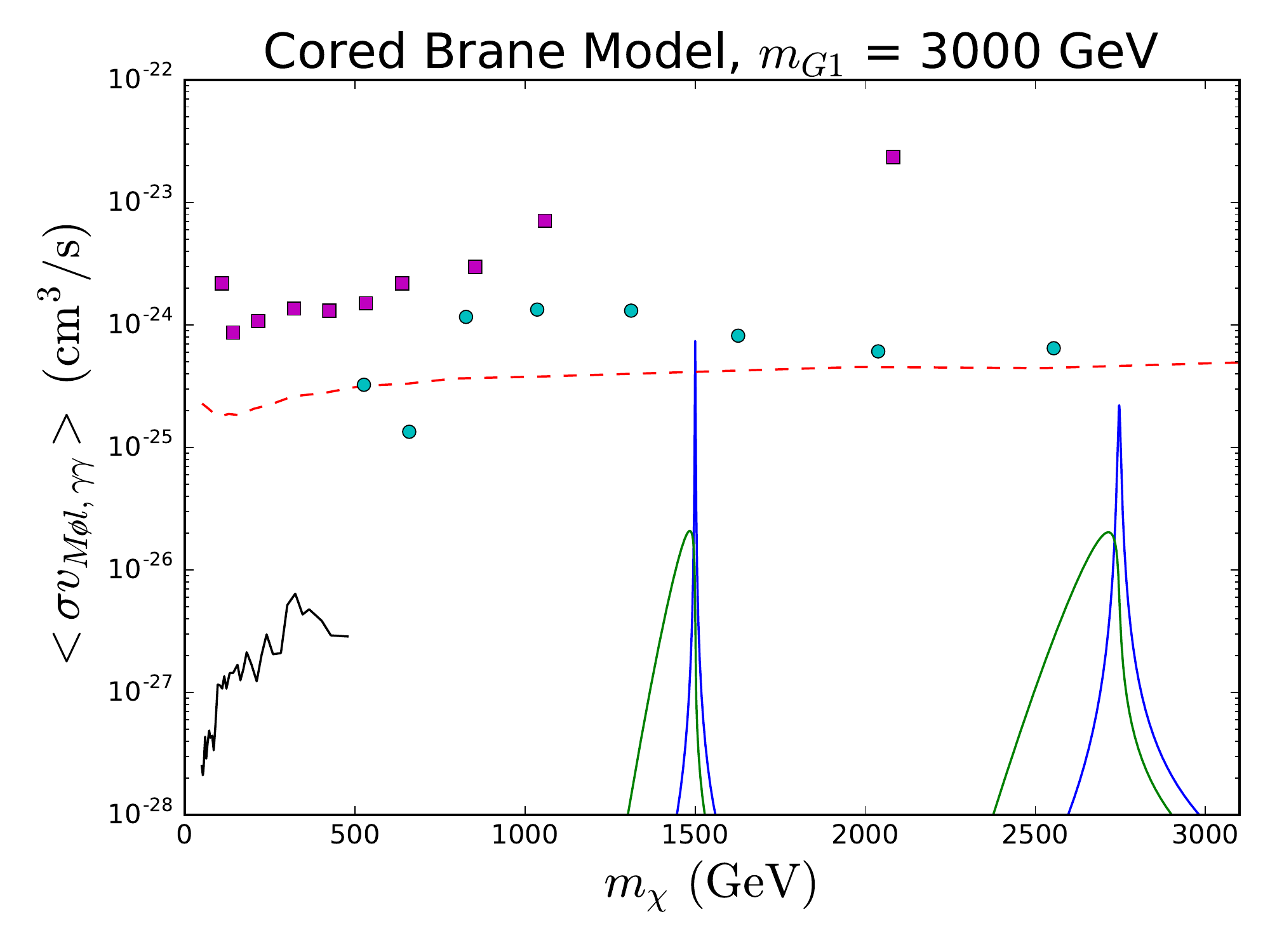}&\includegraphics[scale=0.45]{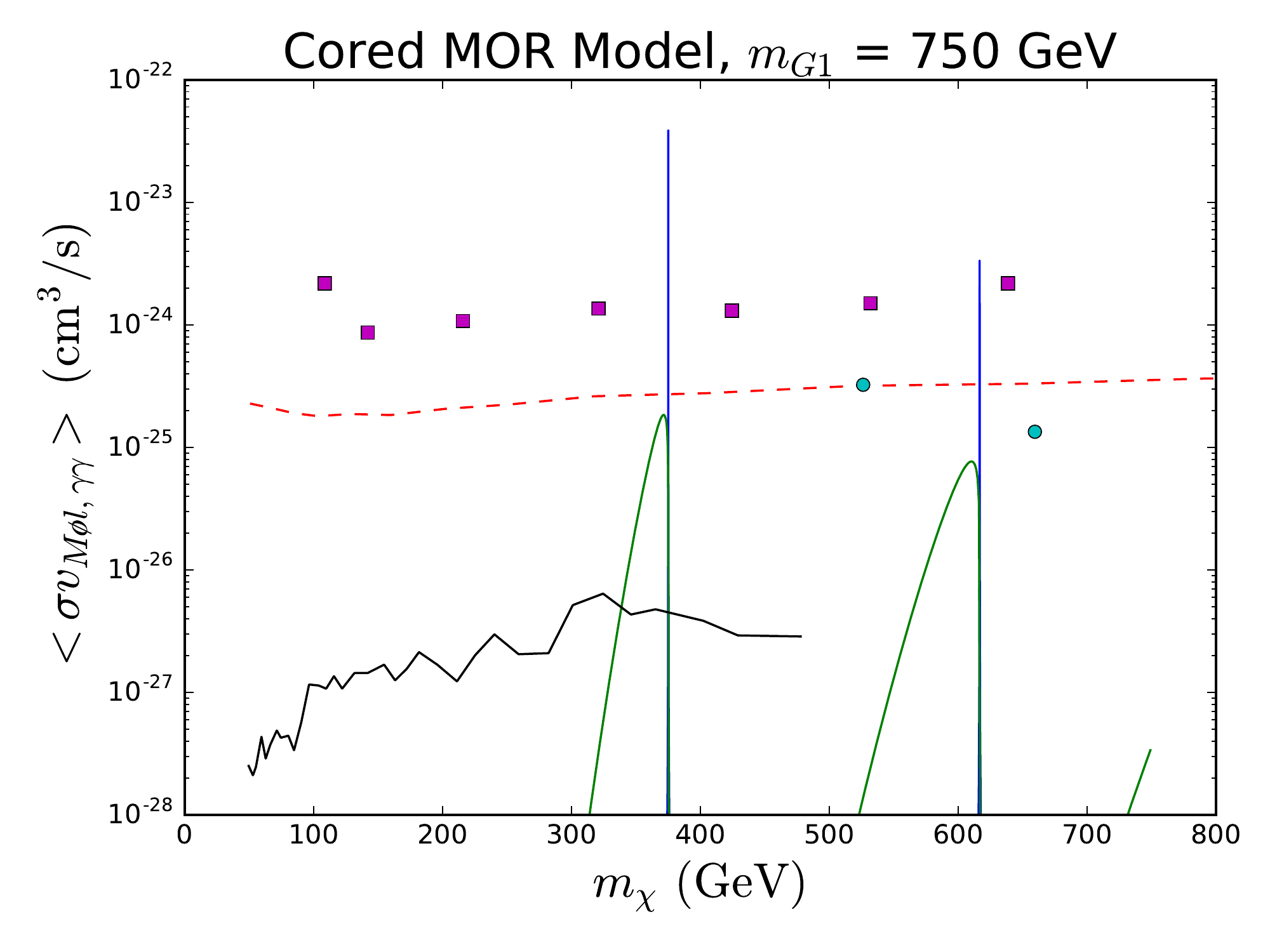}
\end{tabular}

\begin{center}
\vspace{-0.5cm}
\includegraphics[scale=.45]{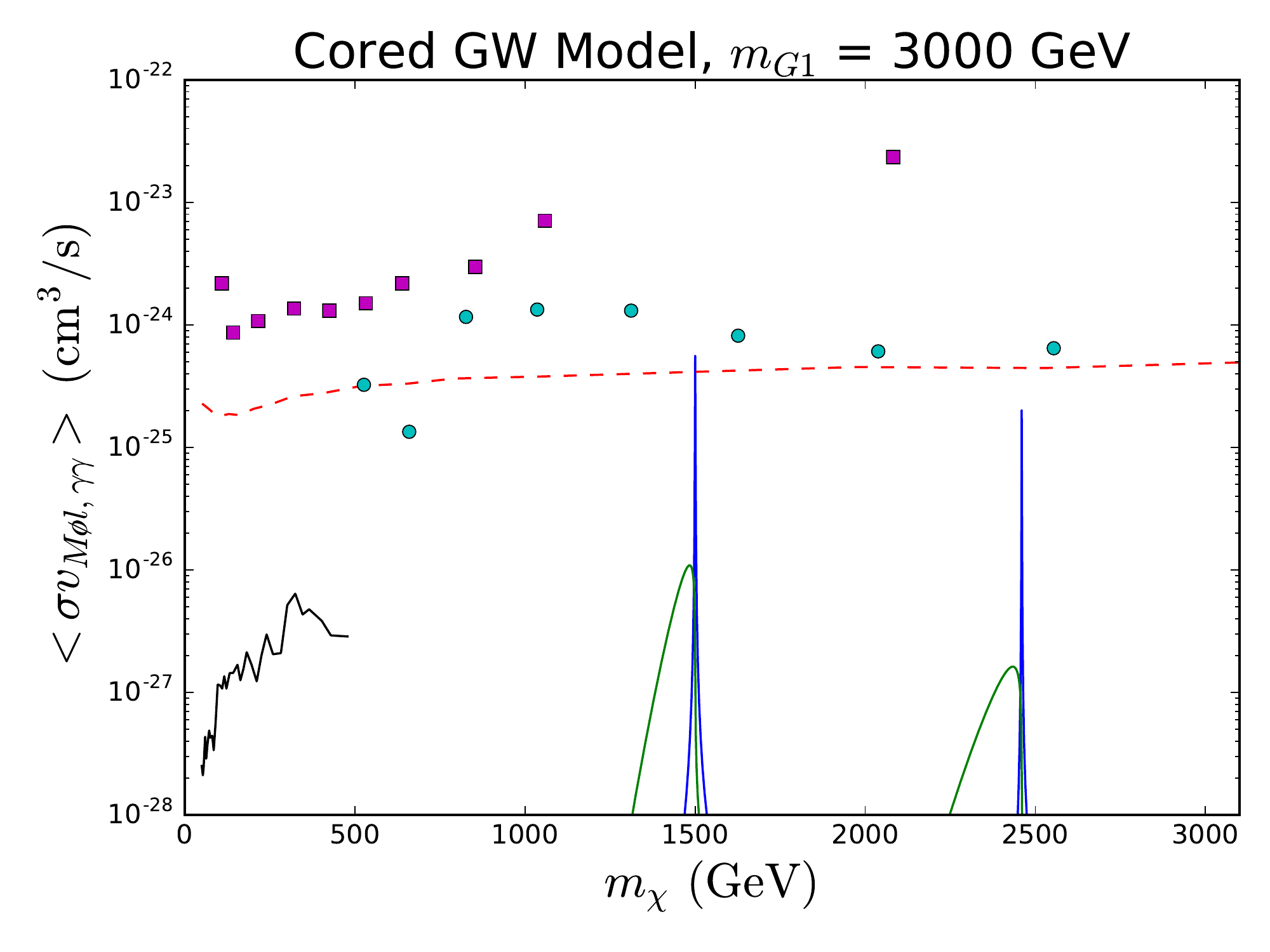}
\end{center}
\vspace{-0.5cm}
\caption{Assuming a cored profile, the thermally averaged dark matter annihilation cross section into two photons at $z=0$
(at thermal freeze-out) for spin-1 dark matter as a function of the DM mass is given by the blue (green) curve for the 5 benchmark models as labeled.
The photon line search limits from FERMI (solid black), HESS I (cyan circles), and 
HESS II (magenta squares) are also shown, as well as the projected 
sensitivity of the Cerenkov Telescope Array (red-dashed line).}
\label{core}
\end{figure}

A second constraint can arise from searches for an excess in the gamma ray continuum from dwarf galaxies; here 
we will make use of the experimental results as presented in Refs.{\cite{Ackermann:2015zua,Drlica-Wagner:2015xua,Ahnen:2016qkx}} 
employing dwarf stacking. The quoted limits assume that the DM annihilates predominantly into a particular SM final state, 
\eg, $\bar bb$, $W^+W^-$ or $\tau^+\tau^-$, which is not the case for the present scenario due to the essentially universal 
nature of the graviton KK couplings.{\footnote {As noted above, the continuum photons, for the non-leptonic annihilation 
modes, are the result of the hadronization of the decay products of the various SM final states.}} As can be seen from 
these references, the $\bar bb$ final state provides the strongest constraint, however, even if it were to be the dominant annihilation 
mode there would be essentially no impact on any of our scenarios. The strongest bound we find is for the case of a 3 TeV 
$G_1$ KK excitation, where in the resonance region the cross section is constrained to be less than $\simeq 2.8 \times 10^{-25}$ 
cm$^3$s$^{-1}$. However, due to the very narrow nature of the DM annihilation resonance peak for $z=0$ essentially none of the parameter space is excluded by this measurement. Of course in the more realistic situation where 
the limit from photon continuum is a weighted combination of the various SM final states, this 
constraint is further weakened by at least a factor of a few in comparison to the full $\bar bb$ channel. Thus we 
conclude that the photon continuum measurements are unlikely to provide constraints on gravity mediated DM annihilation.

A third possible channel which may place constraints on this scenario arises from the bound on the $\bar p$ flux; 
these particles are produced as a result of the decay 
and fragmentation of the SM annihilation products in a manner similar to that of the continuum gamma rays. There are 
two recent theoretical analyses of this bound{\cite {Cuoco:2016eej,Cui:2016ppb}} employing the latest measurements 
from AMS-02{\cite{Aguilar:2016kjl}} which we make use of in our analysis.{\footnote {Perhaps most interestingly, 
both of these theoretical analyses of the AMS-02 data strongly favor an excess consistent with the annihilation of 
$\sim 50-80$ GeV DM into $\bar bb$ or $W^+W^-$ with essentially the thermal cross section reminiscent of the FERMI GC excess.}}   
As in the case of continuum gammas, the theoretical estimate for the $\bar p$ flux is somewhat dependent upon which SM final 
state dominates the DM annihilation process. However, one might expect that most non-leptonic final states will produce 
rather similar bounds; in particular, Ref.\cite {Cuoco:2016eej} shows explicitly that the assumption of dominance of 
either the $\bar bb$ or $W^+W^-$ channels will lead to similar constraints on the $\bar p$ flux. As these authors note
in particular, all of the non-leptonic SM final states yield very similar $\bar p$ spectra and so the constraints they 
yield are robust unless leptonic final states play a significant role in the DM annihilation process. Away from 
the DM signal region, the bounds that these authors obtain on the DM velocity-weighted annihilation 
cross section is a relatively smooth function of the DM mass. Allowing for the systematic uncertainties presented in Fig.7 of 
Ref.{\cite {Cuoco:2016eej}}, and treating the $\bar p$ yield for all of the KK graviton annihilation 
products identically, we find the results for the 5 RS benchmark models as summarized in 
Fig.~\ref{antiprotons}. To obtain these results we included annihilation to all final states except for photons and leptons 
with the approximation that the remaining modes produce similar $\bar p$ spectra. As in the cases above, we display the 
annihilation cross sections for $z=0$, as well as those during freeze-out for comparison. Here we see that for all of the 
benchmark models, as might be expected, the only constraint on the parameter space lies in the narrow graviton 
KK resonance regions. In some cases, \eg, for the GW benchmark, even this very weak constraint is seen to apply only to the 
region near the first KK excitation resonance.  Thus, as was the case for the other final states discussed above, these   
searches provide little if any constraint on RS gravity mediated DM annihilation for the parameter regions discussed here.

\begin{figure}
\begin{tabular}{l l}
\hspace{-1.65cm}\includegraphics[scale=0.45]{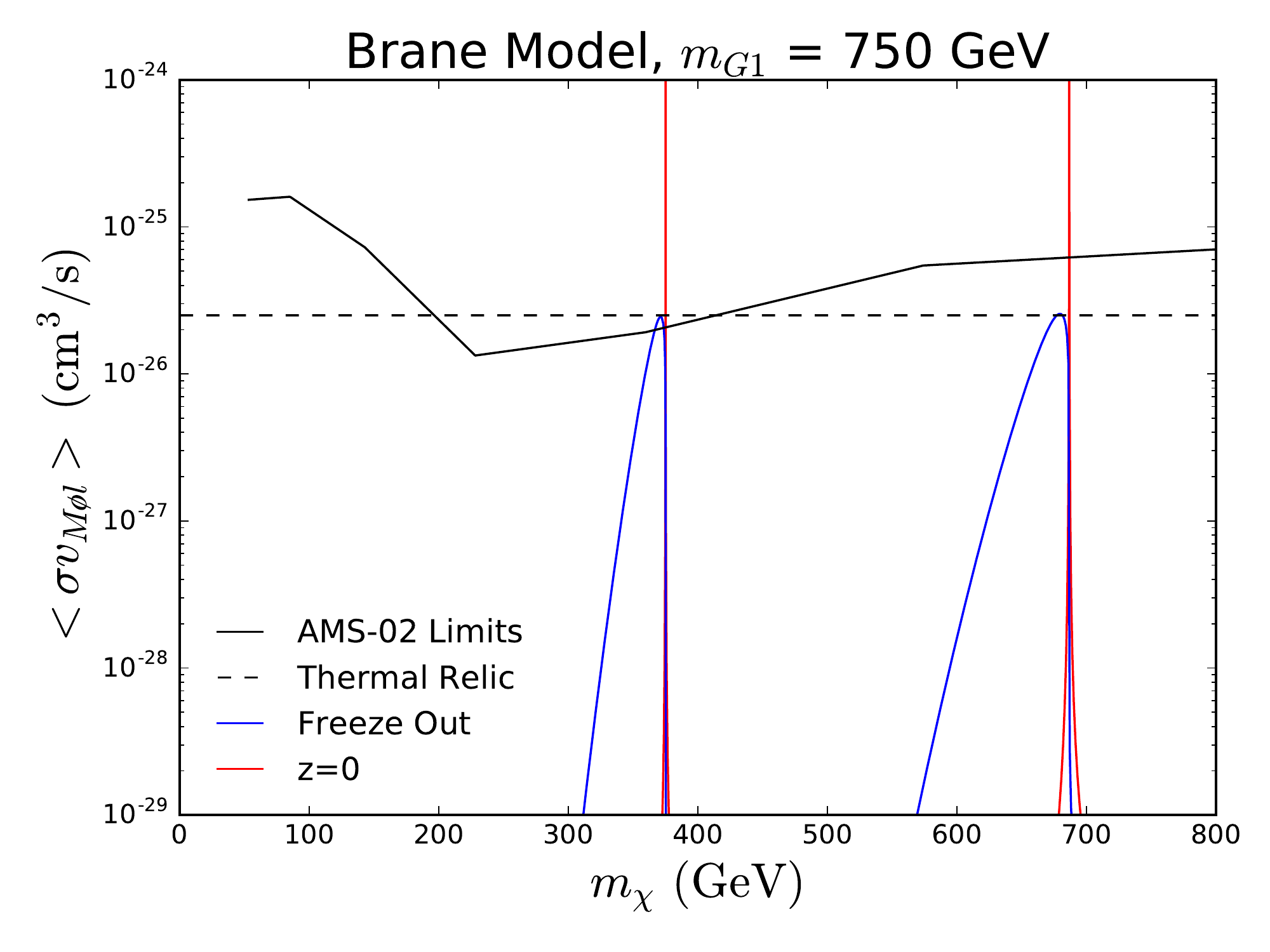}&\includegraphics[scale=0.45]{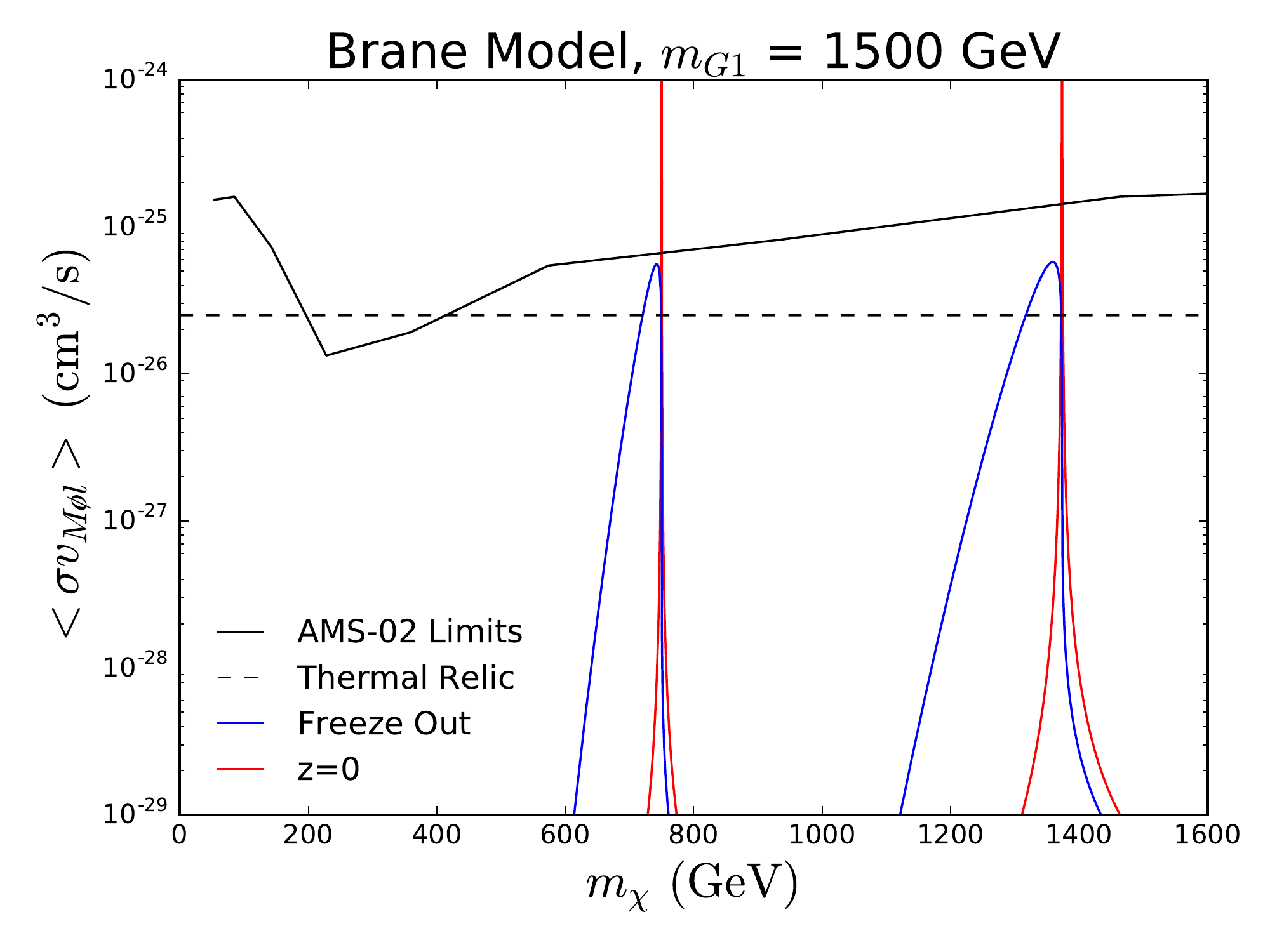} \\
\hspace{-1.65cm}\includegraphics[scale=0.45]{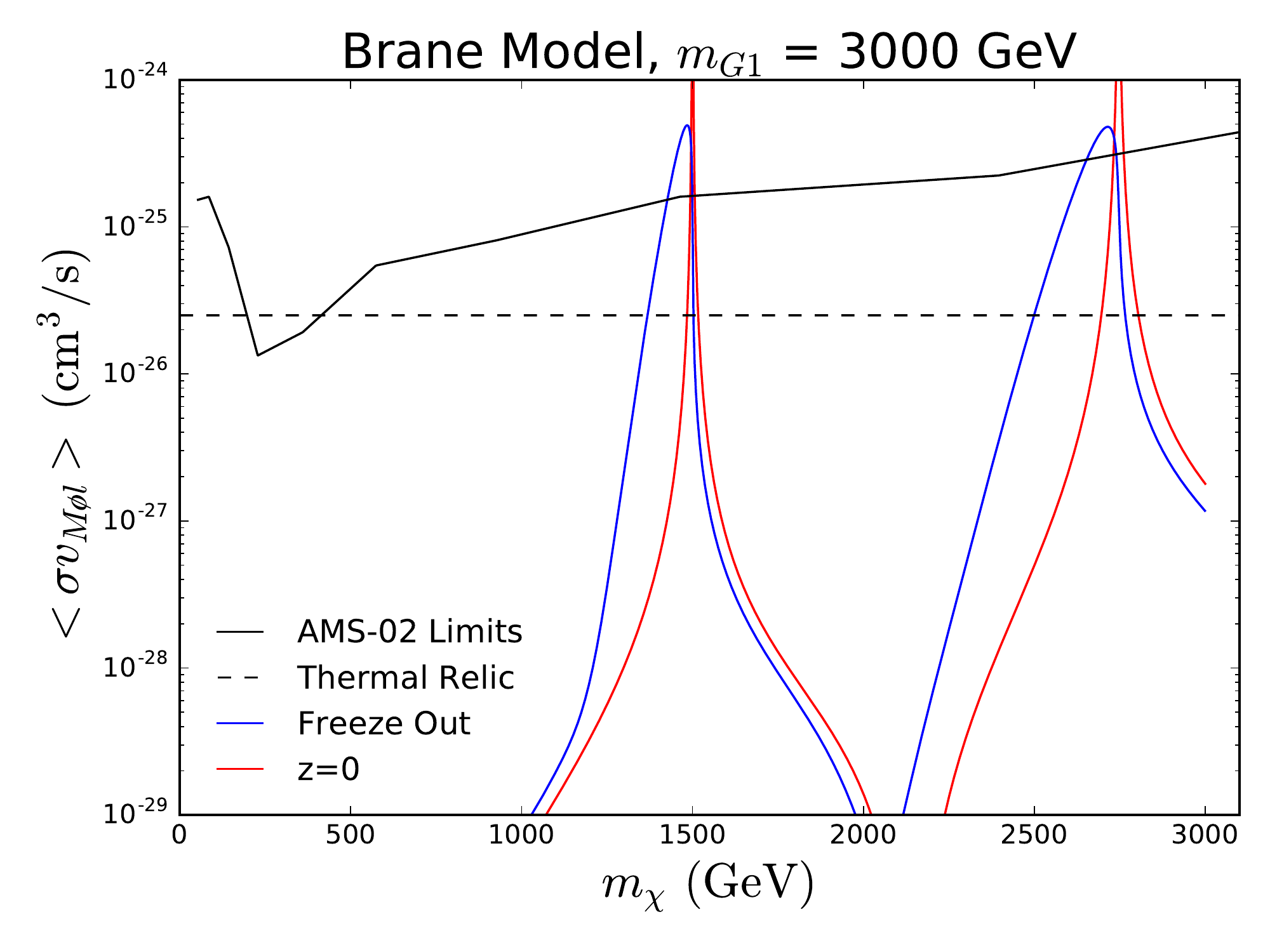}&\includegraphics[scale=0.45]{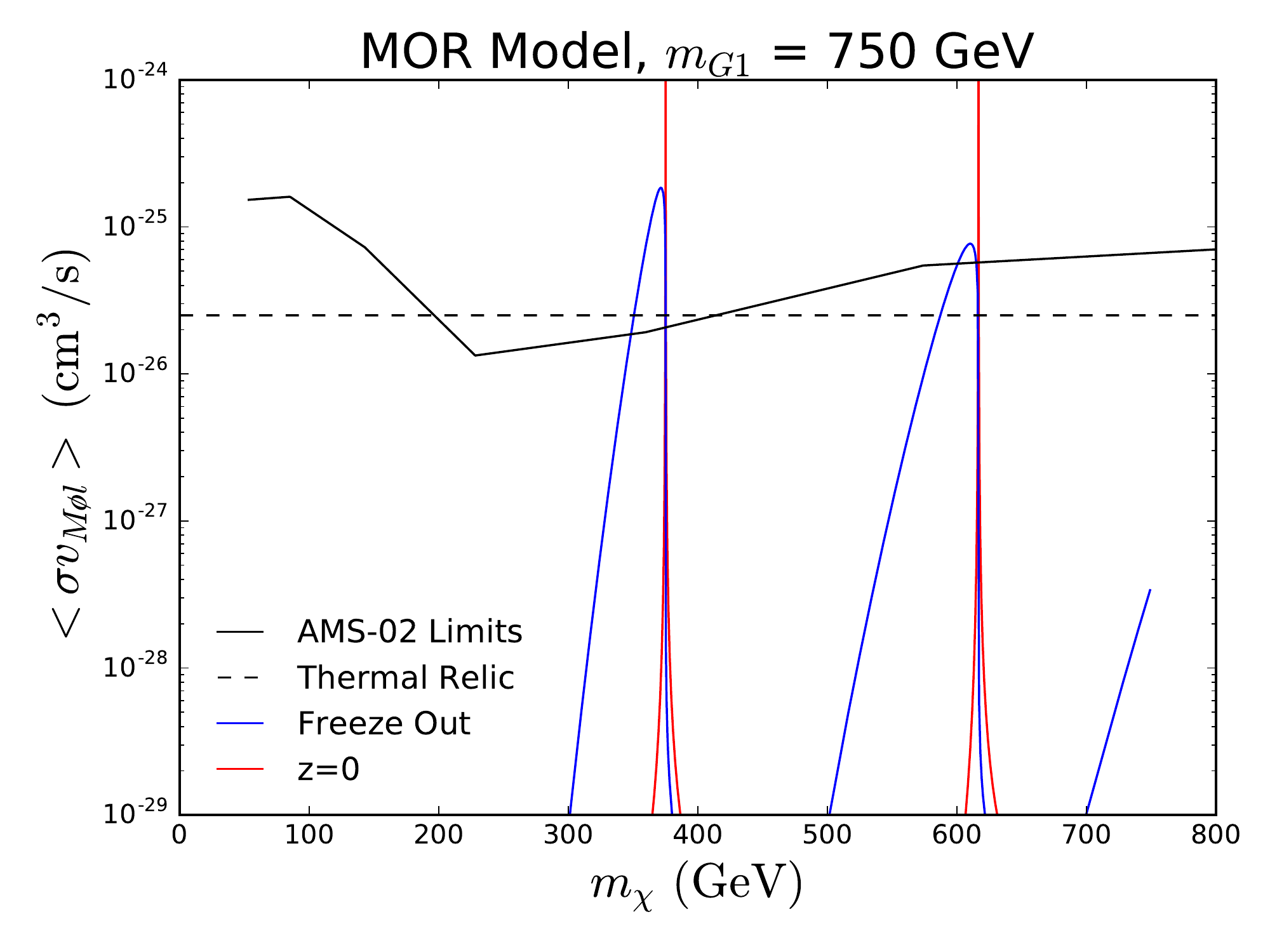}
\end{tabular}
\begin{center}
\vspace{-0.5cm}
\includegraphics[scale=.45]{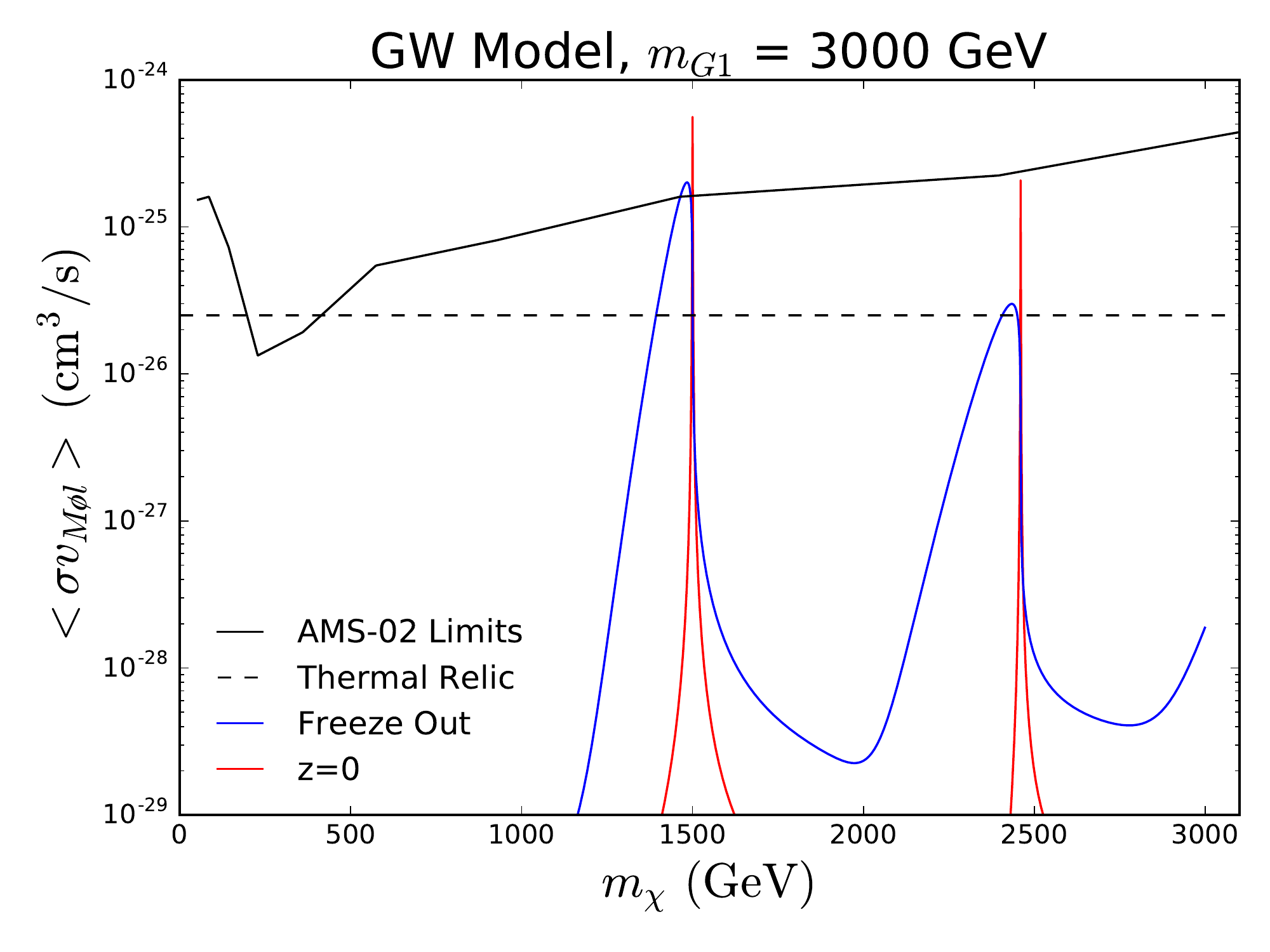}
\end{center}
\vspace{-0.5cm}
\caption{The thermally averaged annihilation cross sections for spin-1 dark matter annihilating into $p \bar{p}$ at z=0 (red) and at freeze-out (blue)
in the 5 benchmark models as a function of the DM mass.  The AMS-02 antiproton constraints correspond to the black curve, while the value of $\left< \sigma v \right>$ that achieves the observed relic abundance is shown as the dashed black curve as a guide for allowed DM masses. }
\label{antiprotons}
\end{figure}

Finally, we consider the possible bounds arising from a potential excess in the flux of positrons due to DM annihilation,
following Ref. \cite{Vittino:2016wjr}, which makes use of the data from AMS-02{\cite {bigams}}. Here, the 
positrons can either be directly produced as DM annihilation products or can be the result of  
decays and fragmentation of the other SM final states. The various constraints for the different annihilation products are summarized
in Ref. \cite{Vittino:2016wjr}. In all cases we see that these constraints 
remain sufficiently loose that none of our benchmark model points lie close to the corresponding bounds even if the channel with the
largest rate is assumed to be applicable. Thus no further constraints on our RS gravity mediated benchmark models are obtained from this
measurement.

\subsection{Dark Matter Spin-Independent Direct Detection}

For completeness, we examine the cross section relevant for direct detection with KK graviton mediation. To begin this discussion we recall the tensor structure of the graviton exchange interaction between two conserved, 
stress-energy sources, $T^{(1,2)}_{\mu\nu}$; schematically we can write this amplitude as proportional to the quantity  
\begin{equation}
{\cal M} \sim T^1_{\mu\nu} ~P^{\mu\nu\alpha\beta}~T^2_{\alpha\beta}\,, 
\end{equation}
where $P$ is the polarization tensor sum that appears in the numerator of the graviton propagator 
\cite{han1999kaluza}. We can rewrite this amplitude in the form (recalling that the stress-energy tensor sources 
are conserved)  
\begin{equation}
{\cal M} \sim  2 ~T^{1\mu\nu} ~T^2_{\mu\nu} - {{2}\over {3}} ~T^1 T^2\,, 
\end{equation}
where $T^i= \eta^{\mu\nu}T^i_{\mu\nu}$ is the trace of the energy-momentum tensor $T^i_{\mu\nu}$. It is convenient to 
recast this amplitude in terms of the trace and traceless components, 
$\tilde T_{\mu\nu}=T_{\mu\nu}-{{1}\over {4}}\eta_{\mu\nu}T$,  of the stress-energy tensor: 
\begin{equation}
{\cal M} \sim 2 ~\tilde T^{1\mu\nu} ~\tilde T^2_{\mu\nu}- {{1}\over {6}} ~T^1 T^2 \,. 
\end{equation}
This form is particularly useful when considering the direct detection (DD) of DM scattering off nuclei, since the nucleon 
matrix elements of both the 
trace and traceless parts of the stress-energy tensor are well-known. For simplicity let us consider first the case where 
all the SM fields lie on the TeV brane. Here, the stress-energy tensor sums over the set of the quark and gluon 
components of the nucleon, all of which couple to the graviton KK tower fields with the same strength. Summing over these 
components yields unambiguous nucleon matrix elements for both the trace and traceless parts of the stress-energy tensor
{\cite{Vecchi:2013iza}}, \ie, $<N|T_N|N>=2m_N^2$ and $<N|\tilde T^N_{\mu\nu}|N>=2(k_\mu k_\nu - \frac{1}{4} \eta_{\mu \nu} m_N^2)$, where $m_N$ is the nucleon mass and $k_\mu$ is the corresponding nucleon 4-momentum. Here we have used the fact that the stress-energy tensor is conserved, 
\ie, $q_{\mu,\nu}T^{\mu\nu}=0$, where $q_\mu$ is the virtual 4-momentum carried by the exchanged KK graviton tower 
between the DM and the nucleus. In order to compute the scattering cross section we take the non-relativistic limit for 
the DM in the lab frame and neglect any small, velocity-dependent corrections. Denoting the KK tower sum as
$m^{-2}_{GR} \equiv \sum_i m^{-2}_i$ where $m_i$ are the $i^{th}$ KK 
tower member graviton masses, we find that the spin-independent (SI) DM-nucleon direct detection cross section for scalar, 
Dirac fermion or vector DM cases discussed above can all be written as 
\begin{equation}
\sigma_N = {{1}\over {9\pi}}~\Big[{{\mu}\over {m_N}}\Big]^2 ~\Bigg[{{m_N} \over {m_{GR}\Lambda_\pi}} \Bigg]^4 ~m_{DM}^2\,, 
\end{equation}
where $\mu$ is the DM-nucleon reduced mass of order $m_N$ for DM masses of interest to us. Note that since we neglect the velocity-dependent corrections and average over the DM spin states, $\sigma_N$ is independent of the DM spin. Recalling that 
DM masses in the vicinity of $\simeq m_i/2$ (for some value of $i$) are needed to saturate the relic density, we find that in 
all cases $\sigma_N$ lies very far below the current LUX/PandaX search limits {\cite {Akerib:2016vxi,Tan:2016zwf}} for any 
interesting values of the parameters, \eg
\begin{equation}
\sigma_N \simeq 3.3\cdot 10^{-54}{\rm cm^2} ~\Big[{m_{DM}\over {500 {\rm GeV}}}\Big]^2 \Bigg[{{30 ~{\rm TeV}^2} 
\over {m_{GR}\Lambda_\pi}} \Bigg]^4 \,. 
\end{equation}
Furthermore, these cross sections are sufficiently small that they are not likely to be probed by direct detection anytime 
in the near future with 
any of the planned experiments. Unfortunately, in the cases where the KK gravitons effectively decouple from the light quarks 
and couple only to the gluons in the nucleon, as in both the GW and MOR benchmarks, these results will be smaller by roughly 
an additional order of magnitude. 



\section{Discussion and Conclusions}

\label{section:Discussion and Conclusions}

In this paper we have examined the possibility that WIMP-like dark matter interacts solely through gravitational interactions,
with the observed relic density being achieved via annihilation into SM fields as mediated by Kaluza-Klein graviton excitations 
within the context of the Randall-Sundrum model. Five benchmark scenarios within this framework were considered in detail
corresponding to various values for the mass of the first graviton KK excitation,
differing SM matter localizations and distinct values of their brane localized kinetic terms. The DM candidate was assumed to 
be a SM singlet with the following three choices of spin being considered:  
A real scalar ($s=0$), a vector-like neutral Dirac fermion ($s=1/2$) or 
a real $U(1)$ massive vector gauge field ($s=1$).   The annihilation cross sections and signals for both indirect and direct detection
were studied in detail.

Independently of the nature of the SM final state, the 
velocity weighted annihilation cross sections for all scenarios were found to scale as $\sim v^{4(1-s)}$, 
where $v$ is the DM velocity in the DM center of mass frame. Since the DM is essentially non-relativistic at freeze-out implying
small velocities, the annihilation of vector DM is then less velocity suppressed and naturally yields a substantially larger cross section than do either scalar or fermionic DM.  This scaling with the DM velocity reflects the 
d(p,s)-wave behavior of the annihilation for the various spin possibilities given the spin-2 nature of the intermediate graviton 
KK excitations. For all benchmark models, vector DM is the only scenario that was found to be capable of reaching 
velocity-weighted annihilation cross sections in the range required to obtain the observed relic density for thermal DM. 
However, even in this case, the required annihilation rate is only obtained in the general neighborhood of one of the graviton KK resonances, 
\ie, for $m_{DM} \simeq 0.5 m_{G_n}$.  We note, however, that for a somewhat more massive first graviton KK state, for example
$\sim 4-5$ TeV, a window of opportunity to generate the observed relic density opens for the case of thermal fermionic DM.  

Direct detection searches for DM in the spin-independent channel (as is relevant for this scenario) were found to be essentially
insensitive to this framework with both current and expected sensitivities being very far from the predicted direct detection cross
section.  The main reason for this small cross section is that the effective operator  
generated by the graviton KK resonances is of dimension-8 since both the couplings of the SM and DM fields to these gravitons 
are non-renormalizable. 

On the other hand, indirect searches for DM could potentially have an impact on small regions of 
the RS parameter space where larger annihilation rates can be obtained.  This region occurs for spin-1 DM with masses
extremely close, \ie, within a graviton KK width, of half the graviton KK mass $m_{DM}=m_{G_n}/2$. Unlike in the case of 
thermal relic annihilation at freeze-out, where Doppler broadening 
widens the effective widths of the KK resonance peaks while decreasing their heights, present day velocity-weighted 
annihilation cross sections (which occur with essentially $v_{DM}=0$)  lead to resonance peaks with widths that 
are determined only by the couplings of the KK graviton resonances themselves.  Thus while a significant range of DM masses can 
generate the observed relic density, only very narrow DM mass windows are subject to present-day indirect searches. 
For these very narrow parameter regions searches for, \eg, gamma ray lines can be of potential importance since the diphoton 
decay branching fraction of our benchmark model points always exceeds $\sim 4\%$. 
However, the anticipated fluxes of these photons are 
quite sensitive to the nature of the DM distribution at the galactic center, and for the case of a cored DM profile even these 
very weak constraints near the KK resonances can be further diminished. Even granting this uncertainty in the DM 
galactic center distribution, we can conclude that photon line searches do not significantly impact this scenario outside of  
extremely narrow DM mass regions. In a similar manner, we find that
searches for DM induced excesses in either the gamma ray continuum or 
for the positron flux are found to yield no additional constraints. Other possible constraints arising from searches for a DM induced 
excess in the $\bar p$ flux can potentially be as, or more, important than in the case of photon line searches. 
However, using the results from 
two recent theoretical analyses of the AMS-02 data we found that, except in the regions very close to the graviton 
KK resonances, these constraints are again found to have essentially no impact on the RS parameter space.  

The Randall-Sundrum framework provides a natural mechanism for thermal DM annihilation to achieve the observed relic density purely through
gravitational interactions.  DM detection is difficult in this scenario, and its best 
test remains the collider production of graviton KK resonances.

\section*{Acknowledgements}

The work was supported by the Department of Energy, Contract DE-AC02-76SF00515.

\newpage

\appendix

\section*{Appendix: Cross sections and Width for Graviton-Mediated Annihilation}

\section{Graviton Width}
\label{appwidth}

We begin by giving expressions for the KK graviton partial widths into scalars, fermions, and vectors (with the appropriate 
suppression factors $\delta$ for the transversely polarized fields which live in the bulk). Note the dark matter always lives 
on the TeV brane except for the GW benchmark point where it is localized similar to the SM $Z$ or $t_R$ depending upon its 
spin. The width into real scalars and fermions, respectively, are
\begin{equation}
\Gamma_{ss} = \frac{\lambda^2 m_G^3}{960 \pi \Lambda_\pi^2} \left(1-4 r_s \right)^{\frac{5}{2}}\,,
\end{equation}
\begin{equation}
\Gamma_{f\bar{f}} = N_c \frac{\lambda^2 m_G^3}{160 \pi \Lambda_\pi^2} \left(1-4r_f \right)^{\frac{3}{2}} \left[ \frac{10}{3}r_f + \frac{(c^f_L)^2+(c^f_R)^2}{2}\left(1-\frac{2}{3} r_f \right) \right]\,,
\end{equation}
where $r_x = m_x^2 / m_G^2$ and $N_c$ is the usual color factor. 
$\lambda$=1 for models with all SM fields on the IR-brane and $\simeq 0.126$ for the MOR benchmark 
with bulk fields. The factors $c^f_{L,R}$ are described in 
the text above and account for the localizations of the left- or right-handed SM fermions in the bulk. For the massless gauge 
bosons, we find the partial width to be
\begin{equation}
\Gamma_{gg, \gamma \gamma} = N_c \delta^2 \frac{\lambda^2 m_G^3}{80 \pi \Lambda_\pi^2}\,,
\end{equation}
where $N_c=8$ for $gg$ and 1 for photons. $\delta=1$ for gauge bosons which live on the IR-brane, 1/2 for gauge bosons 
which live in the bulk for the MOR benchmark, while in the GW scenario one has $\lambda \delta \simeq 1$. 
For the massive SM gauge bosons, we find the corresponding partial widths:
\begin{equation}
\Gamma_{VV} = N \frac{\lambda^2 m_G^3 \sqrt{1-4 r_V}}{480 \pi \Lambda_\pi^2} \left(A_V +  \delta B_V + \delta^2 C_V \right)\,, 
\end{equation}
where
\begin{equation}
\begin{aligned}
& A_V = 1 + 12 r_V + 56 r_V^2\,, \\
& B_V = 80 r_V (1-r_V)\,, \\
& C_V = 12 (1 - 3 r_V + 6 r_V^2)\,.
\end{aligned}
\end{equation}
Here N=1 for distinguishable particles ($V=W$) and 1/2 for identical particles ($V=Z$) in the final state. Note that we 
keep the Higgs on the TeV 
brane in all models considered, so that the Goldstone modes of the massive gauge bosons will not pick up factors of $\delta$.

\section{Dark Matter Annihilation Cross Sections}
\label{xsecapp}
\subsection{Scalar Dark Matter}

Here we consider the differential cross sections for annihilation of DM into the various SM final states. These 
processes are all mediated by the $s$-channel exchange of the massive KK graviton states. We begin with the case of scalar 
DM (denoted here by S). It is useful to define the quantity $P_{ij}$, which arises due to interference between the 
$i^{th}$ and $j^{th}$ KK mode exchanges,
\begin{equation}
P_{ij} = \frac{(s-m_i^2)(s-m_j^2) + m_i \Gamma_i m_j \Gamma_j}{[ (s-m_i^2)^2 + m_i^2 \Gamma_i^2 ] [  (s-m_j^2)^2 + m_j^2 \Gamma_j^2 ]}\,,
\end{equation}
where $m_i$ and $\Gamma_i$ are the mass and width of the $i^{th}$ graviton KK mode, respectively. The first cross 
section we consider is the annihilation of scalar DM into a pair of Higgs bosons ($h$),
\begin{equation}
\frac{d\sigma_{SS\rightarrow hh}}{dz} = \frac{s^3 \beta_S^3 \beta_h^5 (1-3z^2)^2}{9216 \pi \Lambda_\pi^4} \sum_{i,j} \lambda_i^2 \lambda_j^2 P_{ij}\,,
\label{scalsec}
\end{equation}
where as usual $\beta_x = \sqrt{1-4m_x^2/s}$ and $z = cos(\theta)$, with $\theta$ being the scattering angle relative to the 
momentum axis of the incoming particles in the center of mass frame. In a similar fashion we find the differential cross 
section for $SS \rightarrow f \bar{f}$ to be
\begin{equation}
\begin{aligned}
\frac{d\sigma_{SS\rightarrow f \bar{f}}}{dz} = & N_c \frac{s^3 \beta_S^3 \beta_f^3}{4608 \pi \Lambda_\pi^4} \times \left( \sum_{i,j} \lambda_i^2 \lambda_j^2 P_{ij} \right) \times \\
& \left[(1-\beta_f^2)(1+3z^2) + \frac{c^f_{L,i}c^f_{L,j} +c^f_{R,i}c^f_{R,j}}{2}\left( 1-\beta_f^2 +3z^2(1+5\beta_f^2) -18z^4\beta_f^2 \right) \right]\,,
\end{aligned}
\end{equation}
where the first term in brackets corresponds to the contribution from the Dirac mass term (arising from the Higgs vev on the 
IR-brane) and the second term corresponds to bulk left- and right-handed fermion fields. Here $N_c$ is the number of 
colors of the fermion, \ie, =1(3) for leptons(quarks). When the SM Dirac fermion is restricted to the TeV 
brane, $c^f_L = c^f_R = 1$. For the case of neutrinos on the TeV brane, we take $\beta_f =c^f_L=1$ and $c^f_R=0$. For photons and 
gluons in the final state, we find the differential cross-section to be
\begin{equation}
\frac{d\sigma_{SS\rightarrow gg,\gamma \gamma}}{dz} = N \frac{s^3 \beta_S^3 (1-z^2)^2}{512 \pi \Lambda_\pi^4}  \sum_{i,j} \delta_i \delta_j \lambda_i^2 \lambda_j^2 P_{ij}\,.
\end{equation}
For $SS \rightarrow VV$ with massive SM gauge boson final states, we instead obtain
\begin{equation}
\frac{d\sigma_{SS\rightarrow VV}}{dz} = N \frac{ s^3 \beta_S^3 \beta_V}{4608 \pi \Lambda_\pi^4} \sum_{i,j} \lambda_i^2 \lambda_j^2 P_{ij} (A_{SV} + \frac{1}{2}(\delta_i + \delta_j) B_{SV}  + \delta_i \delta_j C_{SV}) \,,
\end{equation}
where
\begin{equation*}
\begin{aligned}
& A_{SV} = 6 + 2 \beta_V^2 (3z^2 -5) + \beta_V^4(5-12z^2+9z^4)\,, \\
& B_{SV} = 4 (1 - \beta_V^2) [3 + \beta_V^2 (2-3z^2)]\,, \\
& C_{SV} = 6 [1 + \beta_V^2(1-3z^2) + \beta_V^4 (1-3z^2 +3z^4)]\,,
\end{aligned}
\end{equation*}
and where as before $N=1$ for $WW$ and 1/2 for $ZZ$. Note that $A_{SV}$ corresponds to the portion of the cross 
section proportional 
to the mass of the gauge boson generated via spontaneous symmetry breaking on the IR-brane, $C_{SV}$ corresponds to the 
cross section resulting from the pure 5-D Yang-Mills kinetic term, and $B_{SV}$ represents the interference between these 
two contributions.

\subsection{Fermionic Dark Matter}

For the case of fermionic DM (denoted by $F$), 
we similarly compute the spin-averaged differential cross sections into the various SM final states. We begin with the result for $F \bar{F} \rightarrow hh$
\begin{equation}
\frac{d\sigma_{F \bar{F} \rightarrow hh}}{dz} = \frac{ s^3 \beta_F \beta_h^5}{18432 \pi \Lambda_\pi^4} (1-\beta_F^2 + 3z^2 (1+2 \beta_F^2) -9z^4 \beta_F^2) \sum_{i,j} \lambda_i^2 \lambda_j^2 P_{ij}\,.
\label{fermsec}
\end{equation}
Next we consider the process $F \bar{F} \rightarrow f \bar{f}$ for SM fermions $f$
\begin{equation}
\frac{d\sigma_{F \bar{F} \rightarrow f \bar{f}}}{dz} =  N_c \frac{s^3 \beta_F \beta_f^3}{36864 \pi \Lambda_\pi^4} \sum_{i,j} \left(A_f + \frac{c^f_{L,i}c^f_{L,j} +c^f_{R,i}c^f_{R,j}}{2} B_f \right) \lambda_i^2 \lambda_j^2 P_{ij}\,,
\end{equation}
where $A_f$ is generated from the Dirac mass term due to the Higgs on the IR-brane and $B_f$ arises from the left- and right-handed SM fields which may propagate in the bulk. $N_c$ is the same color factor as above. $A_f$ and $B_f$ are given by
\begin{equation}
A_f = (1- \beta_f^2)(13 - 4\beta_F^2 +3z^2(7-4\beta_F^2))\,,
\end{equation}
\begin{equation}
B_f = 13 - 4\beta_F^2 + \beta_f^2(5+4\beta_F^2) +3z^2(7-4\beta_F^2 -\beta_f^2(1+20\beta_F^2)) +72z^4 \beta_F^2 \beta_f^2\,.
\end{equation}
For the massless gauge boson final states, we find
\begin{equation}
\frac{d\sigma_{F \bar{F} \rightarrow gg, \gamma \gamma}}{dz} = N \frac{ s^3 \beta_F}{1024 \pi \Lambda_\pi^4} (1-z^2)(2 - \beta_F^2(1-z^2)) \sum_{i,j} \delta_i \delta_j \lambda_i^2 \lambda_j^2 P_{ij}\,,
\end{equation}
where here $N = 8$ for $gg$ and $N=1$ for $\gamma \gamma$. For the massive SM gauge bosons we instead find
\begin{equation}
\frac{d\sigma_{F \bar{F} \rightarrow VV}}{dz} = N \frac{s^3 \beta_F \beta_V}{18432 \pi \Lambda_\pi^4} \sum_{i,j} \lambda_i^2 \lambda_j^2 P_{ij} (A_{FV} + \frac{1}{2}(\delta_i + \delta_j) B_{FV} + \delta_i \delta_j C_{FV})\,,
\end{equation}
with
\begin{equation*}
\begin{aligned}
& A_{FV} = 6(5-2 \beta_F^2) - \beta_V^2 (47 - 20 \beta_F^2 + 3z^2 (4\beta_F^2-7)) + \\
& \hspace{1.1cm} \beta_V^4 (19 - 10\beta_F^2 + 3z^2(8\beta_F^2-5) - 18z^4 \beta_F^2)\,, \\
& B_{FV} = 2 (1-\beta_V^2) [30 - 12\beta_F^2 + \beta_V^2 (17 - 8\beta_F^2 +3z^2(4\beta_F^2-7))]\,, \\
& C_{FV} = 3 \{ 10 - 4\beta_F^2 + \beta_V^2 (3z^2-1)(4\beta_F^2-7) + \\
& \hspace{1.1cm} \beta_V^4 (7 - 4\beta_F^2 -3z^2(1-4\beta_F^2) -12z^4\beta_F^2) \}\,, 
\end{aligned}
\end{equation*}
where N=1 for $WW$ and N=1/2 for $ZZ$ final states as before.

\subsection{Vector Dark Matter}

We now consider the spin-averaged differential cross sections for massive spin-1 DM which lives on the IR-brane. For the GW 
benchmark scenario with DM in the bulk, these expressions are easily adapted. 
We denote the DM here by $X$ (to avoid possible confusion with the massive SM gauge fields $V=W,Z$) and find the differential 
cross section for $XX \rightarrow hh$
\begin{equation}
\frac{d\sigma_{XX \rightarrow hh}}{dz} = \frac{s^3 \beta_h^5 (24 - 8 \beta_X^2 (1+3z^2) +3 \beta_X^4 (1-3z^2)^2 )}{82944 \pi \beta_X \Lambda_\pi^4} \sum_{i,j} \lambda_i^2 \lambda_j^2 P_{ij}\,.
\label{vecsec}
\end{equation}
For final states containing SM fermions we find
\begin{equation}
\frac{d\sigma_{XX \rightarrow f \bar{f}}}{dz} = N_c \frac{s^3 \beta_f^3}{41472 \pi \beta_X \Lambda_\pi^4} \sum_{i,j} \left(C_f + \frac{c^f_{L,i}c^f_{L,j}+c^f_{R,i}c^f_{R,j}}{2} D_f \right)  \lambda_i^2 \lambda_j^2 P_{ij}\,,
\end{equation}
where $C_f$ is generated from the Dirac mass term due to the Higgs vev on the TeV brane and $D_f$ arises from left- and right-handed fields which may propagate in the bulk. $C_f$ and $D_f$ are given by 
\begin{equation}
C_f = (1-\beta_f^2)\left[60 -2\beta_X^2(13+21z^2) +3\beta_X^4(1+3z^2)\right] \,,
\end{equation}
\begin{equation}
D_f = 60 - \beta_X^2(26+42z^2) +3 \beta_X^4(1+3z^2) + \beta_f^2 \left[12 - \beta_X^2(10-6z^2) - \beta_X^4 (3-45z^2+54z^4) \right]\,.
\end{equation}
For the massless gauge boson final states we find
\begin{equation}
\frac{d\sigma_{XX \rightarrow gg, \gamma \gamma}}{dz} = N \frac{ s^3 (16 - 16\beta_X^2(1-z^2) +3\beta_X^4(1-z^2)^2)}{4608 \pi \beta_X \Lambda_\pi^4} \sum_{i,j} \delta_i \delta_j \lambda_i^2 \lambda_j^2 P_{ij}\,,
\end{equation}
and for the massive SM gauge bosons we obtain
\begin{equation}
\frac{d\sigma_{XX \rightarrow VV}}{dz} = N \frac{s^3 \beta_V}{41472 \pi \beta_X \Lambda_\pi^4}  \sum_{i,j} \lambda_i^2 \lambda_j^2 P_{ij} (A_{XV} + \frac{1}{2}(\delta_i + \delta_j) B_{XV} + \delta_i \delta_j C_{XV})\,,
\end{equation}
with
\begin{equation*}
\begin{aligned}
& A_{XV} = 6(30 - 20 \beta_X^2 +3\beta_X^4) - 2\beta_V^2[120 - \beta_X^2 (94-42z^2) + 3\beta_X^4(5-3z^4)] + \\
& \hspace{1.1cm} \beta_V^4 [84 - \beta_X^2(76-60z^2) +3\beta_X^4(5-12z^2+9z^4)]\,, \\
& B_{XV} = 4 (1-\beta_V^2) [90 - 60\beta_X^2 + 9\beta_X^4 + \beta_V^2(30 -\beta_X^2(34-42z^2) + \beta_X^4(6-9z^2) )]\,, \\
& C_{XV} = 6 \{ 30 - 20\beta_X^2 + 3\beta_X^4 + \beta_X^2 \beta_V^2 (3z^2-1)(14-3\beta_X^2) +  \\
& \hspace{1.1cm} \beta_V^4 (18 - 2 \beta_X^2(7-3z^2) + \beta_X^4(3 - 9z^2 +9z^4)) \}\,, 
\end{aligned}
\end{equation*}
with N=1 for $WW$ and N=1/2 for $ZZ$ as before.

\label{app:Xsections}



\end{document}